\documentclass[twocolumn,tighten,trackchanges]{aastex62}

\def\lum{erg~s$^{-1}$}
\def\flux{erg~cm$^{-2}$~s$^{-1}$}

\def\chandra{{\itshape Chandra}}
\def\galex{{\itshape GALEX}}

\def\rosat{{\it ROSAT}}

\newcommand{\nustar}{\textit{NuSTAR}}
\def\suzaku{{\it Suzaku}}
\def\sax{{\it BeppoSAX}}

\newcommand{\xmm}{\textit{XMM-Newton}}
\newcommand{\nswift}{\textit{Neil Gehrels Swift Observatory}}
\newcommand{\swift}{\textit{Swift}}

\def\xray{\hbox{X-ray}}
\def\etal{{et\,al.}}

\def\ltsima{$\; \buildrel < \over \sim \;$}
\def\simlt{\lower.5ex\hbox{\ltsima}}
\def\gtsima{$\; \buildrel > \over \sim \;$}
\def\simgt{\lower.5ex\hbox{\gtsima}}
\def\kms{\ifmmode{~{\rm km~s^{-1}}}\else{~km s$^{-1}$}\fi}
\def\lsim{\lower0.3em\hbox{$\,\buildrel <\over\sim\,$}}
\def\gsim{\lower0.3em\hbox{$\,\buildrel >\over\sim\,$}}
\usepackage{underscore}
\shorttitle{X-ray Spectral Study of ULX M33 X-8}
\shortauthors{West et al.}
\begin{document}

\title{On the Nature of the X-ray Emission from the Ultraluminous X-ray Source,
M33 X-8: New Constraints from \nustar\ and \xmm}

\correspondingauthor{Lacey West}
\email{lad012@uark.edu}

\author{Lacey A. West}
\affil{Department of Physics, University of Arkansas, 226 Physics Building, 825 West Dickson Street, Fayetteville, AR 72701, USA}

\author{Bret D. Lehmer}
\affil{Department of Physics, University of Arkansas, 226 Physics Building, 825 West Dickson Street, Fayetteville, AR 72701, USA}

\author{Daniel Wik}
\affiliation{Department of Physics and Astronomy, University of Utah, 201 James Fletcher Bldg., Salt Lake City, UT 84112, USA}

\author{Jun Yang}
\affiliation{Department of Physics and Astronomy, University of Utah, 201 James Fletcher Bldg., Salt Lake City, UT 84112, USA}

\author{Dominic J. Walton}
\affiliation{Institute of Astronomy, University of Cambridge, Madingley Road, Cambridge CB3 0HA, UK}

\author{Vallia Antoniou}
\affiliation{Department of Physics \& Astronomy, Texas Tech University, Box 41051, Lubbock, TX 79409-€"1051, USA}
\affiliation{Harvard-Smithsonian Center for Astrophysics, 60 Garden Street, Cambridge, MA 02138, USA}

\author{Frank Haberl}
\affiliation{Max-Planck-Institut f\"ur extraterrestrische Physik, Giessenbachstra{\ss}e, 85748 Garching, Germany}

\author{Ann Hornschemeier}
\affil{Laboratory for X-ray Astrophysics, Code 662, NASA Goddard Space Flight Center, Greenbelt, MD 20771, USA}
%\affiliation{NASA Goddard Space Flight Center, Code 662, Greenbelt, MD 20771, USA}

\author{Thomas J. Maccarone}
\affiliation{Department of Physics \& Astronomy, Texas Tech University, Box 41051, Lubbock, TX 79409-€"1051, USA}

\author{Paul P. Plucinsky}
\affiliation{Harvard-Smithsonian Center for Astrophysics, 60 Garden Street, Cambridge, MA 02138, USA}

\author{Andrew Ptak}
\affil{Laboratory for X-ray Astrophysics, Code 662, NASA Goddard Space Flight Center, Greenbelt, MD 20771, USA}
%\affiliation{NASA Goddard Space Flight Center, Code 662, Greenbelt, MD 20771, USA}

\author{Benjamin F. Williams}
\affiliation{Department of Astronomy, Box 351580, University of Washington, Seattle, WA 98195, USA}

\author{Neven Vulic}
\affil{Laboratory for X-ray Astrophysics, Code 662, NASA Goddard Space Flight Center, Greenbelt, MD 20771, USA}
\affil{Department of Astronomy and Center for Space Science and Technology (CRESST), University of Maryland, College Park, MD 20742-2421, USA}
%\affiliation{NASA Goddard Space Flight Center, Code 662, Greenbelt, MD 20771, USA}

\author{Mihoko Yukita}
\affiliation{The Johns Hopkins University, Homewood Campus, Baltimore, MD 21218, USA}
\affil{Laboratory for X-ray Astrophysics, Code 662, NASA Goddard Space Flight Center, Greenbelt, MD 20771, USA}
%\affiliation{NASA Goddard Space Flight Center, Code 662, Greenbelt, MD 20771, USA}

\author{Andreas Zezas}
\affiliation{Harvard-Smithsonian Center for Astrophysics, 60 Garden Street, Cambridge, MA 02138, USA}
\affiliation{Foundation for Research and Technology-Hellas, 100 Nikolaou Plastira Street, 71110 Heraklion, Crete, Greece}
\affiliation{Physics Department \& Institute of Theoretical \& Computational Physics, P.O. Box 2208, 71003 Heraklion, Crete, Greece}



%%%%%%%%%%%%%%%%% ABSTRACT %%%%%%%%%%%%%%%%%%%%%%%
\begin{abstract} 
%250 word limit for the abstract 

We present nearly simultaneous \nustar\ and \xmm\ observations of the nearby
(832~kpc) ultraluminous \xray\ source (ULX) M33 X-8.  M33 X-8 has a 0.3--10~keV luminosity of $L_{\rm X}
\approx 1.4 \times 10^{39}$~\lum, near the boundary of the ``ultraluminous''
classification, making it an important source for understanding the link between
typical Galactic \xray\ binaries and ULXs.  Past studies have shown that the
0.3--10~keV spectrum of X-8 can be characterized using an advection-dominated
accretion disk model.
%M33 X-8 is one of the nearest ``broadened disk'' ULXs, the 0.3--10~keV spectra
%of which are characterized well by an advection dominated accretion disk.  In
%the context of the broader ULX population, BD ULXs typically have \xray\
%luminosities that are near the boundary of typical sub-Eddington black hole
%\xray\ binaries and super-Eddington ULXs.  
We find that when fitting to our \nustar\ and \xmm\ observations, an additional
high-energy ($\simgt$10~keV) Comptonization component is required, which allows
us to rule out single advection-dominated disk and classical sub-Eddington
models.  With our new constraints, we analyze \xmm\ data taken over the last 17
years to show that small ($\approx$30\%) variations in the 0.3--10~keV flux of
M33 X-8 result in spectral changes similar to those observed for other ULXs.
%At low fluxes, the spectrum is characterized well using an advection-dominated
%disk at low energies ($\simlt$3~keV) with a power-law-like component at high
%energies ($E \approx$~3--10~keV).  As the flux rises, the high-energy component
%shows increasing curvature, which we suggest is likely due to either (1)
%cooling and truncation of the accretion disk and corresponding increase in the
%importance of a cool and more optically thick Comptonization component; or (2) an
%increase in luminosity of the advection-dominated disk.  These scenarios are
The two most likely phenomenological scenarios suggested by the data are degenerate in terms of constraining the nature of the accreting compact object (i.e., black hole versus neutron star). We further present a search for
pulsations using our suite of data; however, no clear pulsations are detected.
%We argue that even if M33 X-8 contains a pulsating NS, the signatures would
%be diluted by the powerful accretion disk emission.
%Such behavior is consistent with recent ULX funnel models that feature
%outflowing material from the innermost regions of an advection dominated disk.
%The rising flux increases the accretion disk radius by which material is
%ejected and thus the region obscured by the Comptonizing component.  The flux
%and spectral variations of X-8 are small and constraints based on 0.3--10~keV
%data are relatively weak compared to our \xmm\ and \nustar\ campaign.  
Future observations designed to observe M33 X-8 at different flux levels across
the full 0.3--30~keV range would significantly improve our constraints on the
nature of this important source.  
%\color{red}XXX---THIS NEEDS TO BE SHORTENED---XXX \color{black}

\end{abstract}



%%%%%%%%%%%%% Introduction %%%%%%%%%%%%%%%%%
\section{Introduction} \label{sec:intro}

%\latex\ \footnote{\url{http://www.latex-project.org/}} is a document markup
% \citep{lamport94}%\aastex\ 
%\footnote{see Section \ref{sec:pubcharge} in the Appendix for more details}

Ultraluminous \xray\ sources (ULXs) are often defined as off-nuclear \xray\
point sources with luminosities that exceed $L_{\rm X} \approx 10^{39}$~\lum,
the classical Eddington limit for a 10~$M_\odot$ black hole (BH).  Over the
last two decades, new observations and theoretical models suggest that most
ULXs are predominantly powered by super-Eddington accretion onto neutron stars
(NSs) and stellar-mass BHs (i.e., $\simlt$10~$M_\odot$), as opposed to
sub-Eddington accretion onto intermediate-mass BHs (see Kaaret \etal\ 2017 for
a comprehensive review).  Much of these insights have been gained thanks to
\xmm\ and \nustar\ observations of the brightest nearby ULXs (see, e.g.,
Gladstone \etal\ 2009; Walton \etal\ 2018a), and the emergence of theoretical
models that have been instrumental in explaining the observed spectra.  

Modulo some uncertainty on details, it is generally thought that
as accretion onto the compact object approaches the Eddington limit, radiation
pressure will increase the scale height of the innermost portions of the disk,
leading to local mass loss in a radiatively-driven wind and inward advection of 
energy, as per the ``slim'' disk model (e.g., Abramowicz \etal\ 1988; Mineshige
\etal\ 2000).  The bulge of the inner disk and wind columns can form a
funnel-like structure that leads to geometric beaming of radiation for vantage
points that are close to the rotation axis (e.g., Poutanen \etal\ 2007;
King~2009; Dotan \& Shaviv 2011).  For the case of a NS accretor, the accretion
disk may be interrupted within the magnetospheric radius, where material will
flow along the magnetic field lines to the NS magnetic poles (e.g., King \&
Lasota~2016; Mushtukov \etal\ 2017).  

Whether NS or BH accretors dominate the ULX population as a whole is a subject
of current debate.  At present, there are only four extragalactic ULXs that
have been discovered to contain NS accretors on the basis of pulsations:
M82~X-2, NGC 7793 P13, NGC~5907 ULX1, NGC 300 ULX1, and SMC X-3 (e.g., Bachetti \etal\
2014; F{\"u}rst \etal\ 2016; Israel \etal\ 2017; Carpano \etal\ 2018; Tsygankov \etal\ 2017), and also
one Galactic accreting pulsar ULX: \swift\ J0243.6+6124 (Wilson-Hodge \etal\
2018).  There are also a few Galactic BH sources that reach super-Eddington
accretion rates, such as GRS~1915+105 (e.g., Done \etal\ 2004), V4641 Sgr
(Revnivtsev \etal\ 2002), and V404 Cyg (Jourdain \etal\ 2017), suggesting ULXs
are very likely to be a mixed NS and BH population.  However, population
synthesis arguments (see, e.g., Wiktorowicz \etal\ 2017) and observational
biases against detecting pulsations (e.g., King \etal\ 2017, Mushtukov \etal\
2017) suggest that many more ULXs may contain NS compact objects than
previously thought.  Recent observational studies have shown that the hard
spectral shape associated directly with the pulsed emission in pulsar ULXs can
be used to successfully model the hard spectral component of non-pulsar ULXs,
suggesting that many more NS ULXs may be lurking in the broader ULX
population (e.g., Walton \etal\ 2018b, hereafter, W18; Koliopanos \etal\ 2017;
Pintore \etal\ 2017).

%Recently, Walton \etal\ (2018b; hereafter, W18) showed that broad band (i.e.,
%0.3--30~keV) pulsar spectra can be modeled well using three component models,
%in which an inner advection dominated disk and outer standard disk contribute
%two thermal components and a higher-energy Comptonization component, which
%often dominates above 10~keV, in the regime probed only by \nustar.  Using
%phase-resolved spectra, they showed that the pulsed components of the ULXs 
%provide the Comptonization, which they associate with the accretion column that
%is guided onto the NS poles within the magnetospheric radius.  They estimate
%that such accretion columns contribute $>$50\% of the 0.3--40~keV fluxes of
%ULXs with detectable pulsars.  When analyzing {\it all} ULXs with high
%signal-to-noise broad band spectra available (including \xmm\ or \suzaku\ and
%\nustar), they find similarly shaped hard spectral components are required in
%their fits, suggesting some (or perhaps all) of the non-pulsar ULXs may in fact
%be NSs (see also Pintore \etal\ 2017).  Furthermore, the majority of ULXs without known pulsars show that the
%hard component is generally a smaller fraction of the overall 0.3--40~keV flux,
%explaining why pulsed emission has not been observed in many ULXs.  

To date, the majority of the broad band high signal-to-noise ULX spectra have
come from highly super-Eddington objects, with $L_{\rm X} = $(5--100)~$\times
10^{40}$~\lum.  In this paper, we explore the case of M33 X-8, a relatively
nearby ($D = 832$~kpc; Bhardwaj \etal\ 2016), low-luminosity ($L_{\rm
0.3-10~keV} \approx 10^{39}$~\lum) ULX.  Given its corresponding high flux
($10^{-11}$~\flux), which is comparable to the fluxes of many of the luminous
ULXs, M33 X-8 provides the best opportunity to constrain the properties of a
ULX that is at the transition between luminous Galactic BH XRBs and the broader
extragalactic ULX population.  The source is located in the nuclear region of
M33, which initially suggested that the X-8 was a possible AGN (Long \etal\
1981).  However, a lack of any expected luminous optical counterparts (Long
\etal\ 2002) and tight upper limits on the central supermassive black hole mass of
M33 ($\simlt$1500~$M_\odot$; e.g., Gebhardt \etal\ 2001; Davis \etal\ 2017),
%and the observed 106~day modulated \xray\ period of the source (Dubus \etal\
%1997) 
indicated that the source is a ULX.  Furthermore, there are a number of
young ($\simgt$50~Myr) early-type stars in the nuclear region of M33, with
which the source could be associated (Long \etal\ 2002; Garofali \etal\ 2018).

A comprehensive spectral characterization of M33 X-8 was previously performed
by Middleton \etal\ (2011; hereafter, M11) using 12 \xmm\ observations that
spanned a three-year baseline.  M11 performed stacked spectral fitting for
three distinct flux (luminosity) ranges and found that the 0.3--10~keV spectrum
of M33 X-8 could be modeled well using both sub-Eddington (hot accretion disk
plus optically-thin/hot Comptonization) and super-Eddington (slim disk and slim
disk plus optically-thick/cold Comptonization) models, preventing tight constraints on
the nature of the accretion onto this source.

Here, we present a first coordinated broad band (0.3--30~keV) spectrum from
temporally overlapping 24~ks \xmm\ and 102~ks \nustar\ observations, and
provide spectral and timing analyses of this data set ($\S$3.1).  We consider both the
models provided by M11, which were presumed to include a BH accretor, and also
the more recent NS-based model from W18 that provides good fits to the broad
band spectra of luminous ULXs.  Using knowledge of the spectral components from our \xmm\
plus \nustar\ fits, we revisit the spectral variability of this source using
observations from the \xmm\ archive, which includes 17 \xmm\ observations that
span 17 years.  To further assess the nature of the source, we perform a search
for pulsations in this object.  Throughout this paper, we assume a Galactic
column density of $N_{\rm H} = 5.7 \times 10^{20}$~cm$^{-2}$ in the direction
of M33 (Dickey \& Lockman~1990), and include this column in all spectral fits.
Unless stated otherwise, quoted uncertainties throughout this paper correspond
to 90\% confidence intervals.

%
%%%%%%%%%%%%%%%%%%%%%%%%%%%%%%%%%%%%%%%%%%%%%%%%%%%%%%%%%%%%%%%%%%%%%%%%%%%%%%%%%%
% Figure 1
%%%%%%%%%%%%%%%%%%%%%%%%%%%%%%%%%%%%%%%%%%%%%%%%%%%%%%%%%%%%%%%%%%%%%%%%%%%%%%%%%%
%
\begin{figure*}[t!]
\figurenum{1}
\centerline{
\includegraphics[width=19cm]{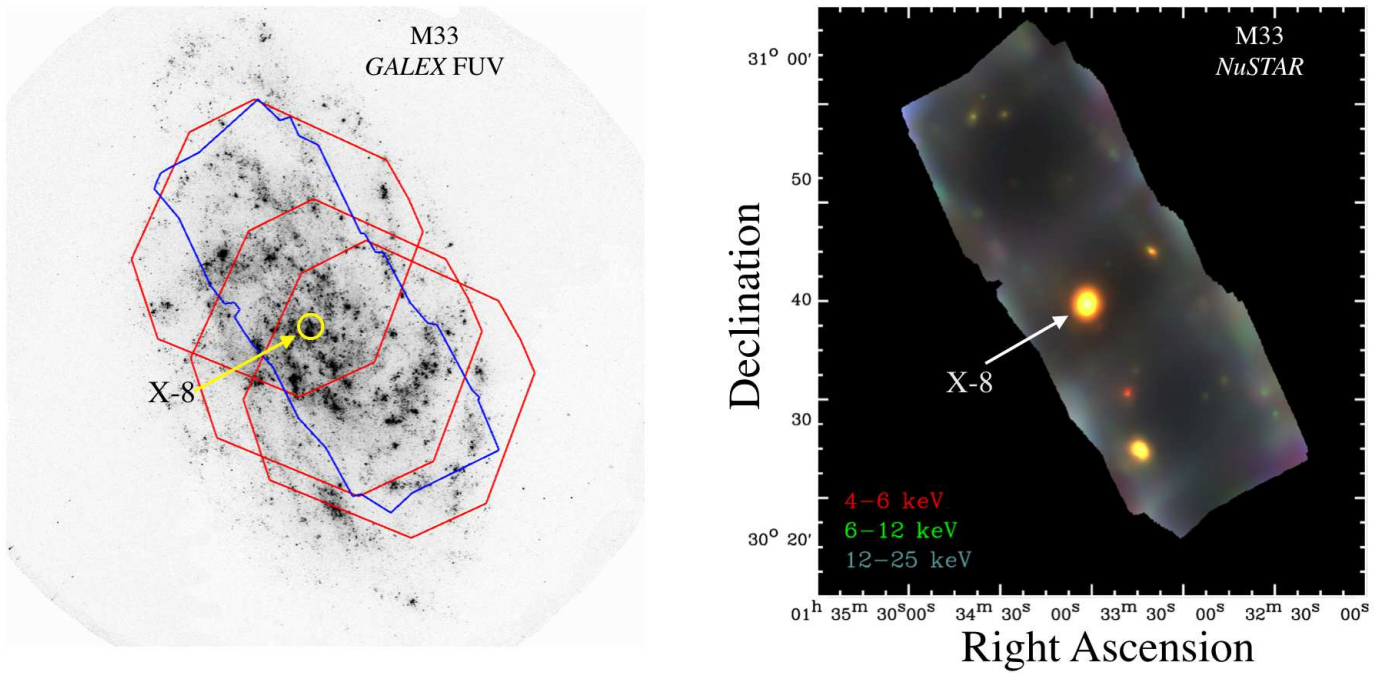}
}
\vspace{-0.1in}
\caption{
%%%
({\it Left\/}) \galex\ FUV image of M33 with the aerial footprints of
the \nustar\ ({\it blue region\/}) and recent \xmm\ ({\it red regions\/})
observations (from July--August 2017) are overlaid.  The position of M33 X-8 is
indicated.  Our observational strategy permitted X-8 to be observed in all
three \xmm\ exposures.  ({\it Right\/}) Three-color \nustar\ image mosaic of
the M33 legacy fields.  The image was constructed from 4--6~keV ($red$),
6--12~keV ($green$), and 12--25~keV ($blue$) exposure-corrected adaptively
smoothed images.  
%%%
}
\end{figure*}
\begin{table*}[t]
\begin{center}
      \caption{\xmm\ Observation Log}
      \label{tab:galaxy}
        \begin{tabular}{l c c c c c c c c} \toprule
          %\multicolumn{1}{c }{} & \multicolumn{7}{c}{\xmm} & \multicolumn{1}{c }{ } & \multicolumn{1}{c}{} & \multicolumn{1}{c}{} \\  
	   &  & \multicolumn{3}{c}{Useful Exposure (ks)} & \multicolumn{1}{c}{} & \multicolumn{1}{c }{} & \multicolumn{1}{c }{ } &  \\
	   &  & \multicolumn{3}{c}{\rule{1.3in}{0.01in}} & \multicolumn{1}{c }{Off-Axis Angle}  & \multicolumn{1}{c }{$F_{\rm X}$ (0.3--10~keV)}  & & \multicolumn{1}{c}{}   \\

	  \multicolumn{1}{c}{Obs Date}  & \multicolumn{1}{c}{Obs ID}  & PN & MOS1 & MOS2 &  (arcmin) & ($10^{-11}$ ergs s$^{-1}$) & Bin & \multicolumn{1}{c}{PI}  \\ 
	  \multicolumn{1}{c}{(1)}  & \multicolumn{1}{c}{(2)}  & (3) & (4) & (5) &  (6) & (7) & (8) & \multicolumn{1}{c}{(9)}  \\ \hline
	  2000 Aug 02 & 0102640401 &  0 & 13.0 & 12.9 & 13.2 & 1.52  & 2 & B. Aschenbach \\
	  2000 Aug 04 & 0102640101$^a$ & 7.5 & 10.6$^c$ & 10.5$^c$ & 0.3 & 1.67 & 3 & B. Aschenbach \\
	  2000 Aug 07 & 0102640301$^a$ & 3.7$^d$ & 10.0 & 7.3 & 13.6 & 1.68 & 3 & B. Aschenbach \\
	  2001 Jul 05 & 0102640601 & 3.2$^{b,d}$  & 4.7$^d$ & 4.7$^d$ & 7.4 & 1.69 & 3 & B. Aschenbach \\
	  2001 Jul 05 & 0102640701 & 0 & 11.5 & 11.6 & 13.2 & 1.73 & 4 & B. Aschenbach \\
	  %2001 July 07 & 0102640801 & 0\tablenotemark{\dag\dag} & 1.8\tablenotemark{\dag\dag} & 1.7\tablenotemark{\dag\dag} & 13.4 & & & B. Aschenbach  \\
	  %2001 July 08 & 0102641001 & 1.3\tablenotemark{\dag}\tablenotemark{\dag\dag} & 9.7\tablenotemark{*}\tablenotemark{\dag\dag} & 9.7\tablenotemark{\dag\dag} & 11.0 & 1.72 & 4 & B. Aschenbach \\
	  2001 Aug 15 & 0102642001$^a$ & 8.8 & 11.5 & 11.5 & 13.3 & 1.81 & 4 & B. Aschenbach \\
	  2002 Jan 25 &  0102642101$^a$ & 10.0$^b$ & 12.3 & 12.3 & 10.6 & 1.67 & 3 & B. Aschenbach \\
	  2002 Jan 27 & 0102642301$^a$ & 10.0$^d$  & 12.3 & 12.3 & 7.2 & 1.56 & 2 & B. Aschenbach \\
	  2003 Jan 23 & 0141980601$^a$ & 11.5 & 13.5 & 13.6 & 13.2 & 1.85 & 4 & W.~Pietsch \\
	  2003 Jan 24 & 0141980401$^a$ & 0 & 1.8 & 1.4 & 13.6 & 1.83 & 4 & W.~Pietsch  \\
	  2003 Feb 12 & 0141980801$^a$ & 8.1 & 10.2 & 10.2 & 0.3 & 1.46 & 1 & W.~Pietsch \\
	  2003 Jul 11 & 0141980101$^a$ & 6.3$^d$  & 7.3 & 8.3 & 11.0 & 1.61 &  3 & W.~Pietsch \\
	  %2003 Jul 25 & 0141980301 & 9.0$^b$ & 11.2\tablenotemark{*} \tablenotemark{**} \tablenotemark{\dag\dag} & 12.3\tablenotemark{**} \tablenotemark{\dag\dag} &  7.4 & & & Pietsch et al. (2004)\\
	  %2003 Jan 22 & 0141980501 & 0.9\tablenotemark{\dag\dag} & 6.4\tablenotemark{*}\tablenotemark{\dag\dag} & 6.8\tablenotemark{\dag\dag} & 0.3 & 1.80 & 4 & Pietsch et al. (2004) \\
	  2010 Jul 09 & 0650510101 & 69.6$^d$  & 100.8 & 100.7$^{b,d}$ & 7.8 & 1.74 & \ldots$^{f}$ & B.~Williams  \\
	  2010 Jul 11 & 0650510201 & 71.0 & 101.1 & 101.1 & 5.2 & 1.54 & 2 & B.~Williams \\
	  %2012 Jan 10 & 0672190301 & 0\tablenotemark{\dag\dag} & 116.1\tablenotemark{**}\tablenotemark{\dag\dag} & 115.2\tablenotemark{**}\tablenotemark{\dag\dag} & 14.1 & & & B.~Williams et al. (2015)\\
	  2017 Jul 21 & 0800350101 & 18.6 & 21.6 & 21.6 & 8.3 & 1.79 & 4 & B. Lehmer \\ 
	  2017 Jul 23 & 0800350201$^e$  & 20.9$^b$ & 22.6 & 22.6 & 1.5 & 1.58 & 2 & B. Lehmer \\   
	  2017 Aug 02 & 0800350301 & 21.6$^{d}$  & 23.3 & 23.3 & 10.9 & 1.62 & 3 & B. Lehmer \\
\hline\hline
      \end{tabular}
\end{center}
Note---The medium optical blocking filter was used for all three EPIC cameras in each of these observations except in ObsIDs: 0102640301 (thin filter 2 used for MOS2), 0102640401 (thick filter used for all three detectors), 0650510101 and 0650510201 (thin filter 1 used for pn in both).
Col.(1): Start date of the observation.  All rows are order by ascending date.  Col.(2): Unique \xmm\ observation ID.
Col.(3)--(5): Cumulative good-time-interval exposure times for pn, MOS1, and MOS2 after filtering for
flaring (see $\S$2.2 for details). Col.(6): EPIC pn-based off-axis angle in
units of arcmin. Col.(7): 0.3--10~keV flux, as measured by fitting the
individual ObsID pn data to a {\ttfamily TBABS $\times$ (DISKPBB + COMPTT)}
model (see $\S$3.2 for details). Col.(8): Designated flux bin value used for
joint spectral fitting in $\S$3.3.  Sources with $F_{\rm X}$/($10^{-11}$~\flux)
$< 1.5$, 1.5--1.6, 1.6--1.7, and $> 1.7$ are given bins 1, 2, 3, and 4,
respectively. Col.(9): PI of the ObsID.  Original works from archival
observations can be found in Pietsch \etal\ (2004) and Williams \etal\ (2015).\\
$^a$ MOS observations used by M11.\\
$^b$ Corrected for pileup following the procedure described in $\S$2.2.\\
$^c$ Exposure taken in Small Window Mode.\\
$^d$ Core of X-8 PSF on or very near (within $\sim25$\arcsec) a CCD gap or FOV edge.\\
$^e$ Observation exposure overlaps NuSTAR observation (ObsID 50310002003).\\
$^f$ Excluded from flux-binned spectral fitting due to high fractional variability and investigated individually in $\S$3.2.\\
     %$^{\rm (1)}$ Angle of offset between \color{red}EPIC \color{black} observation target coordinates and M33 X-8 \\
     %$^{\rm (2)}$ Flux of X-8 based on spectral fits (baseline model; see Section 3.1) to the pn data \\
     %$^{\rm (3)}$ Observations are arbitrarily binned according to source flux, with the limits $F_{\rm X}$/($10^{-11}$~\flux) $< 1.5$, 1.5--1.6, 1.6--1.7, and $> 1.7$ constraining bins 1, 2, 3, and 4, respectively (see Section 3.2 for further discussion). \\
      %$^{\rm ()}$ The medium optical blocking filter was used for all three EPIC cameras in each of these observations except in ObsIDs: 0102640301 (thin filter 2 used for MOS2), 0102640401 (thick filter used for all three detectors), 0650510101 and 0650510201 (thin filter 1 used for pn in both). \\
      %$^{\rm ()}$ \color{red} Reference for all observations before Feb 2003:    Pietsch, W., Misanovic, Z., Haberl, F., et al. 2004, A\&A, 426, 11 \color{black} \\
      %$^{\rm ()}$ \color{red} Reference for B. Williams' observations:      Williams B. F., et al., 2015, ApJS, 218, 9 \color{black}\\
      %$^{\rm ()}$ \color{red} Note: B. Aschenbach's observations were included in Pietsch et al. (2004) but he was not included or cited in that paper - leave his name in PI column for those observations?
     %$^{\rm \dag\dag}$Observation suffered from slew failure, affected by radiation, telemetry glitches, etc. [Orange/red warning text in obs. log]
\end{table*}

%%%%%%%%%%%%% Section 2 %%%%%%%%%%%%%%%
\section{Data and Spectral Extractions} \label{sec:analysis}

M33 was observed by \nustar\ in three unique $12 \times 12$~arcmin$^2$ fields
over two separate epochs (i.e., six total observations) that occurred in
February and March of 2017 (epoch 1; ObsID: 50310002001) and July and August of
2017 (epoch 2; ObsID: 50310002003) as a part of the M33 \nustar\ Legacy
program.\footnote{See https://www.nustar.caltech.edu/page/59 for details on
\nustar\ Legacy programs.}    In addition to placing constraints on the nature
of M33 X-8 (the subject of this paper), the M33 \nustar\ Legacy program was
designed to help characterize accretion states for the XRB population
throughout the galaxy.  A forthcoming publication (Yang \etal\ in-prep) will
address the properties of the M33 XRB population more broadly.  Figure~1 shows
\galex\ FUV and \nustar\ three-color mosaic images of the M33 region.  

The first epoch of \nustar\ observations covered X-8 in one of the three
observations.  Unfortunately, only one nearly simultaneous observation with the
\nswift\ was conducted during this epoch, and it did not contain X-8 in its
field of view (FOV).  We therefore do not make use of \swift\ data in this
study.  Furthermore, given the lack of nearly simultaneous low-energy (i.e.,
$<$3~keV) data, and a brief inspection showing little difference between the
\nustar\ spectra between epochs 1 and 2 (the 3--30~keV spectrum of epoch~1 is
consistent with a very small constant offset of $\approx$2\%), we chose not to
make use of the epoch 1 data when analyzing spectra.  We do, however, consider
the epoch~1 data when searching for pulsations; and details of this analysis
are presented in $\S$3.4.  The second epoch of \nustar\ observations, which
constitute the focus of this study, were accompanied by nearly simultaneous
\xmm\ observations of the three fields.  During epoch~2, X-8 was observed once
by \nustar; however, thanks to the larger field of view (FOV) of \xmm\ ($27.5
\times 27.5$~arcmin$^2$ for pn and $33 \times 33$~arcmin$^2$ for MOS), X-8 was
covered by all three of the \xmm\ observations (see red \xmm\ pn FOV outlines
in Fig.~1).  The new observations, presented here, thus consist of three \xmm\
exposures of M33 X-8 (each 23--25~ks), with one \xmm\ epoch (ObsID: 0800350201) being
nearly simultaneous with the 102~ks \nustar\ ObsID: 50310002003, which occurred
on 2017 July 23.

%
%%%%%%%%%%%%%%%%%%%%%%%%%%%%%%%%%%%%%%%%%%%%%%%%%%%%%%%%%%%%%%%%%%%%%%%%%%%%%%%%%%
% Figure 2
%%%%%%%%%%%%%%%%%%%%%%%%%%%%%%%%%%%%%%%%%%%%%%%%%%%%%%%%%%%%%%%%%%%%%%%%%%%%%%%%%%
%
\begin{figure}
\figurenum{2}
\centerline{
\includegraphics[width=9cm]{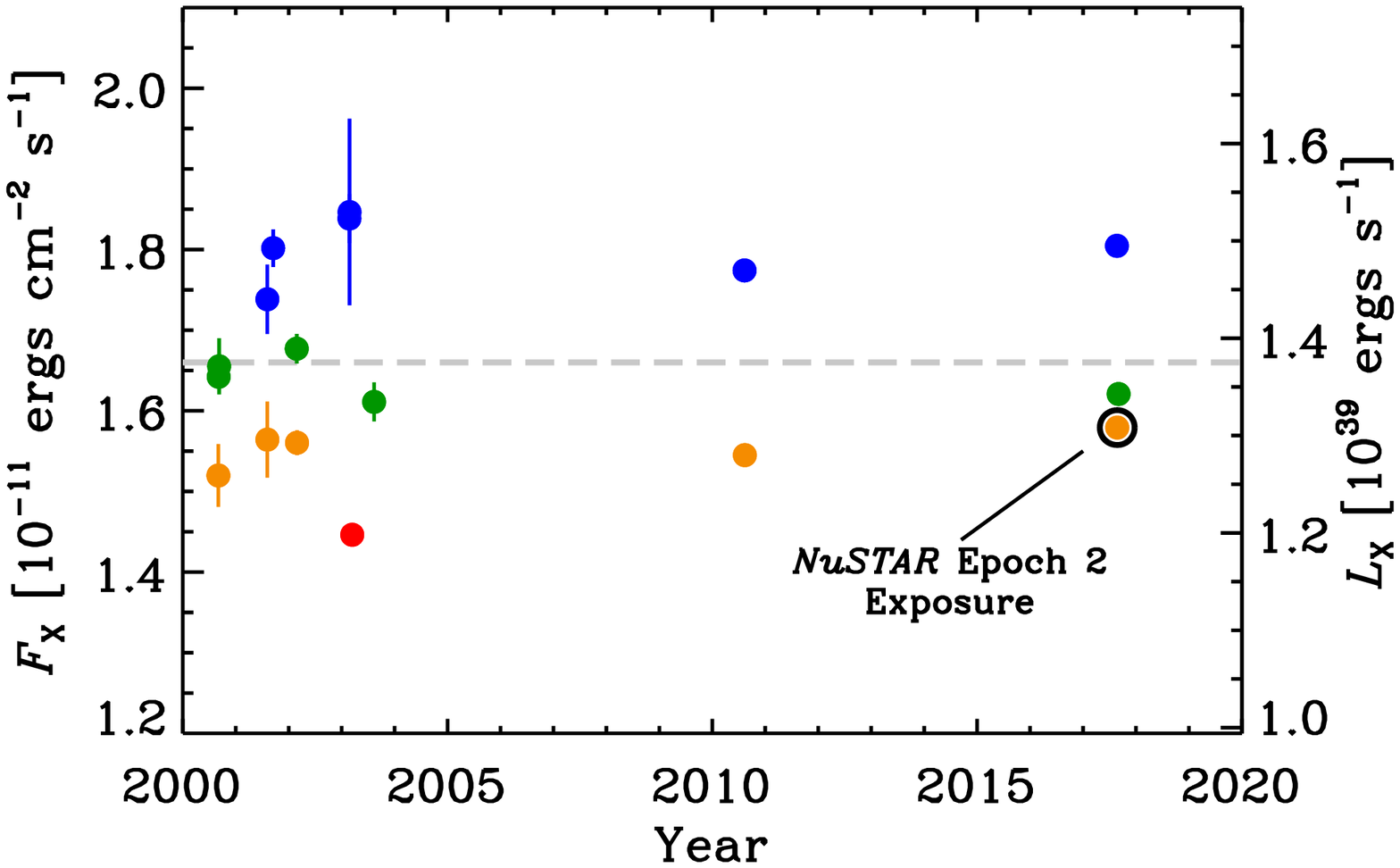}
}
\vspace{0.1in}
\caption{
%%%
0.3--10~keV flux (and luminosity) versus time for the \xmm\ EPIC-pn archival
observations.  Colors represent different flux bins, as used in the spectral
grouping performed in $\S$3.2, with $F_{\rm X}$/($10^{-11}$~\flux) $< 1.5$,
1.5--1.6, 1.6--1.7, and $> 1.7$, displayed as red, orange, green, and blue,
respectively.  The observation containing the nearly simultaneous \nustar\
exposure of M33 X-8 is annotated with a black open circle.  The flux of M33 X-8
over the last 17 years has remained very stable with a mean value and standard
deviation of $F_{\rm X} = (1.66 \pm 0.10) \times 10^{-11}$~\flux\ ($L_{\rm X} =
[1.38 \pm 0.09] \times 10^{39}$~\lum).  The \xray\ flux of X-8 was near its
average value during our simultaneous \xmm\ and \nustar\ observation.  
%%%
}
\end{figure}

When analyzing the spectral properties of X-8, we first use the nearly
simultaneous \nustar\ and \xmm\ observations during epoch~2 to constrain the 0.3--30~keV
spectrum.  We use this observational set to create ``baseline'' spectral
models of the source, which we then apply to both the new \xmm\ data from 2017 (described
previously in this section), as well as archival \xmm\ data.  In Table~1, we provide an \xmm\
observation log for the data used in this program, and in Figure~2, we show the
measured 0.3--10~keV flux of X-8 versus time for the entire \xmm\ archival
history.  Several of the available \xmm\ observations (ObsIDs 0102640801, 0102641001, 0141980301, 0141980501 and 0672190301) that contain M33 X-8 in the FOV suffered from observation issues (e.g., slew failure, high levels of background radiation, and telemetry glitches) and were therefore excluded from this study.
In the subsections below, we describe our data reduction and
spectral extraction procedures pertaining to X-8.

\subsection{\textit{NuSTAR} Reductions}

The \nustar\ data was reduced using {\ttfamily
HEASoft} v6.20, \nustar\ Data Analysis Software ({\ttfamily NuSTARDAS}) v1.7.1,
and CALDB version 20170503.  We processed level 1 data to level 2 products by
running {\ttfamily nupipeline}, which performs a variety of data reduction
steps, including (1) filtering out bad pixels, (2) screening for cosmic rays
and observational intervals when the background was too high (e.g., during
passes through the SAA), and (3) projecting accurately the events to sky
coordinates by determining the optical axis position and correcting for the
dynamic relative offset between the optics bench and focal-plane bench due to
motions of the 10~m mast that connects the two benches.  A total ``cleaned''
exposure of 101.5~ks in epoch~2 was utilized for this analysis.

M33 X-8 source spectra were extracted from a 40\arcsec\ radius circular
aperture, centered at ($\alpha$,$\delta$)$_{\rm J2000}$ = 01$^{\rm h}$~33$^{\rm
m}$~50.6$^{\rm s}$, $+30^\circ$~39\arcmin~31\arcsec, the centroid of the
\nustar\ point-source emission.  The aperture size was chosen to both encompass
a large fraction of the point-spread function (PSF) and also ensure large
signal-to-noise for the highest possible energies in our spectra.  We further
inspected 2--8~keV images from the \chandra\ ACIS Survey of M33 (ChASeM33;
T{\"u}llmann \etal\ 2011) and found that there were no additional sources
within this aperture to the deep limits in ChASeM33.  We extracted background
spectra from three circular apertures, located well outside the PSF of X-8, yet
within the same chip as X-8, in regions without any obvious sources detectable
by eye.  These background regions cover a total area of 17.28~arcmin$^2$, and
were chosen to be the same between FPMA and FPMB.  Source and background
spectra, along with redistribution matrices and auxiliary response files (RMFs
and ARFs) were produced using {\ttfamily nuproducts} with the spectra grouped
to a minimum of 50 counts per energy bin and analyzed across the 3--30 keV
energy range, where the source counts were observed to exceed the background
counts.

\subsection{\textit{XMM-Newton} Reductions}

The reduction of \textit{XMM-Newton} data was carried out with the \xmm\
Science Analysis System (SASv16.0.0), following the procedures recommended in
the online user guides.\footnote{See the
\href{https://xmm-tools.cosmos.esa.int/external/xmm_user_support/documentation/sas_usg/USG/}{\textit{XMM-Newton}
SAS User Guide} and 
\href{https://www.cosmos.esa.int/web/xmm-newton/sas-threads}{SAS threads}
for details.}
Calibrated events lists for the EPIC pn (Str{\"u}der \etal\ 2001)
and MOS (Turner et al.  2001) detectors were produced using {\ttfamily epchain}
and {\ttfamily emchain}, respectively.  Good time intervals (GTIs) were
established from the full field light curve between 10~keV and 12~keV for pn
and at $>10$ keV for MOS; periods of relatively high count rate ($> 0.4$ counts
s$^{-1}$ for pn and $> 0.35$ counts s$^{-1}$ for MOS) were removed from the
events lists to account for background flaring.  Standard filters were then
applied to allow only single and double events ({\ttfamily PATTERN$<$=4} and
{\ttfamily FLAG==0}) in the pn events list and only single events ({\ttfamily
PATTERN==0} and {\ttfamily flag = \#XMMEA\_EM}) for MOS.

Pile-up was evaluated using the ratios of modeled-to-observed single and double
patterns which are output by {\ttfamily epatplot} for a circular region of
radius $45$\arcsec\ around the source in all detectors.  When the ratio of
singles was less than 1 and doubles greater than 1, the observation was
considered to be piled-up.  For such observations, pile-up was corrected by
extracting spectra from annuli with outer radii of 45\arcsec\ and inner radii
of 5\arcsec\ and 10\arcsec\ for MOS and pn, respectively.
%Pileup in the MOS detectors (MOS 2 in the case of ObsID 0650510101) was
%accounted for by excising the core (inner $5$\arcsec\ radius circular region)
%of the PSF when extracting source products from the filtered events list.
%\color{red} In the case of pileup in the pn detector, a new events list was
%created to account for X-ray loading ({\ttfamily epchain keepintermediate=all
%runepxrlcorr=yes}), filtered as outlined above and later used in the creation
%of the RMF to correct for pileup. \color{black} 
Pile-up corrections were implemented for the pn data in ObsIDs 0102640601,
%0102641001, 
0102642101, 
%0141980301 
and 0800350201 and for MOS2 in ObsID
0650510101 following our procedure.  Out-of-time (OoT) events were found to be
negligible ($\simlt$0.3\%) in all observations and were therefore not removed
from pn or MOS spectra.
%\color{red} procedure of
%excising the PSF core (inner $\sim10$\arcsec) from the extraction region.
%\color{red}Out-of-time (OoT) events were found to be negligible in all
%observations and were therefore not removed from pn or MOS spectra.
%\color{black}

RMFs and ARFs were generated using SAS commands {\ttfamily rmfgen} and
{\ttfamily arfgen}, respectively. %In the absence of pileup, Lightcurves and
Source spectra were then extracted from the fully filtered events lists in
circular regions with radii of $45$\arcsec (or annular regions in the case of
pile-up; see above), centered on the source with the SAS command {\ttfamily
evselect}.  Background spectra were extracted from the same events lists, in circular
regions with radii of $45$\arcsec\ on the same CCD as the source but free from obvious
contaminating point sources (i.e., sources visible by eye in 0.3--10~keV
images).  
%For X-8, the back
%Lightcurves were corrected for the background count rate using {\ttfamily
%epiclccorr}.
Spectra were grouped to have a minimum of 50 counts per energy bin, and
analyzed across the full 0.3--10.0 keV energy range.

%%%%%%%%%%%%%%%%%%%%%%%%%%%%%%%%%%%%%%%%%%%%%%%%%%%%%%%%%

%
%%%%%%%%%%%%%%%%%%%%%%%%%%%%%%%%%%%%%%%%%%%%%%%%%%%%%%%%%%%%%%%%%%%%%%%%%%%%%%%%%%
% Figure 3
%%%%%%%%%%%%%%%%%%%%%%%%%%%%%%%%%%%%%%%%%%%%%%%%%%%%%%%%%%%%%%%%%%%%%%%%%%%%%%%%%%
%
\begin{figure}
\figurenum{3}
\centerline{
\includegraphics[width=9cm]{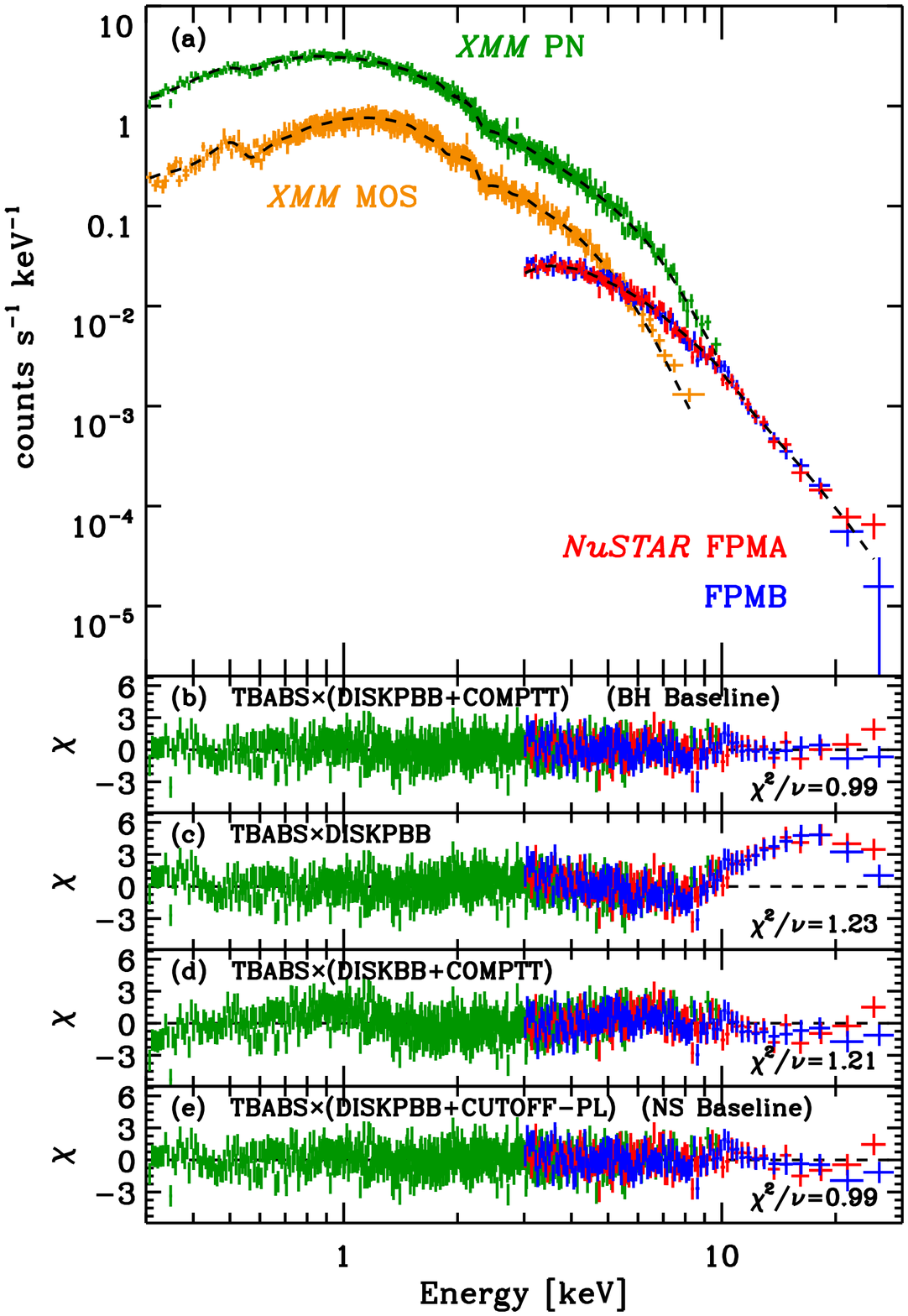}
}
\vspace{0.1in}\caption{
%%%
({\it a}) Best-fit 0.3--30~keV spectrum of M33 X-8 for our nearly-simultaneous
\nustar\ (ObsID: 50310002003) and \xmm\ (ObsID: 0800350201) observations.  The
data are shown for the \xmm\ pn ({\it green}), MOS1 and MOS2 ({\it orange}),
and \nustar\ FPMA ({\it red}) and FPMB ({\it blue}) cameras.
%%%
Best-fit residuals for pn and \nustar\ cameras are shown (MOS is excluded for
clarity of viewing) for models ($b$) {\ttfamily TBABS $\times$ (DISKPBB +
COMPTT)}, which is our BH baseline model, ($c$) {\ttfamily TBABS $\times$
DISKPBB}, ($d$) {\ttfamily TBABS $\times$ (DISKBB + COMPTT)}, and ($e$) our NS
baseline model {\ttfamily TBABS $\times$ (DISKPBB + CUTOFF-PL)}.  Detailed
physical descriptions of these models are provided in $\S$3.1.  The M11 study
of X-8 had shown that models ($b$)--($d$) provided acceptable fits to the \xmm\
archival data; however, we find that only our baseline models are
acceptable when using the combined \xmm\ plus \nustar\ data.
%%%
}
\end{figure}
%%%%%%%%%%%%%%%%%%%%%%%%%%%%%%%%%%%%%%%%%%%%%%%%%%%%%%%%%%%%%%%%%%%%%%%%%%%%%%%%%%

%%%%%%%%%%%%%%%%%%%%%%%%%%%%%%%%%%%%%%%%%%%%%%%%%%%%%%%%%%%%%%%%%%%%%%%%%%%%%%%%%%
% Table 2
%%%%%%%%%%%%%%%%%%%%%%%%%%%%%%%%%%%%%%%%%%%%%%%%%%%%%%%%%%%%%%%%%%%%%%%%%%%%%%%%%%
\begin{table*}
\begin{center}
\caption{Best fit parameters for Nearly-Simultaneous \nustar\ plus \xmm\ Observation.}
\begin{tabular}{lccccc}
\hline\hline
 &  &  &  & (BH Baseline) & (NS Baseline) \\ %
\multicolumn{1}{c}{\sc Parameter} & {\sc Unit} & {\ttfamily DISKPBB} & {\ttfamily DISKBB + COMPTT} & {\ttfamily DISKPBB + COMPTT} & {\ttfamily DISKPBB + CUTOFF-PL} \\ %
\hline
\hline
$N_{\rm H}$ \dotfill                           &   10$^{22}$ cm$^{-2}$     & $0.169 \pm 0.005$           & $<0.0004$               & $0.133_{-0.006}^{+0.016}$   &  $0.142 \pm 0.006$           \\
$kT_{\rm in}$ \dotfill                         &   keV                     & $1.73 \pm 0.03$             & 0.751$ \pm 0.018$       & 1.06$_{-0.22}^{+0.29}$      &  $1.37 \pm 0.03$             \\
$p$ \dotfill                                   &                           & $0.523 \pm 0.004$           & \ldots            & $0.569_{-0.024}^{+0.009}$   &  $0.553_{-0.006}^{+0.007}$   \\
$(R_{\rm in}/D_{10})^2 \cos \theta$ \dotfill   &                           & $0.0267 \pm 0.0027$         & $1.77_{-0.11}^{+0.13}$  & $0.25_{-0.15}^{+0.26}$      &  $0.086_{-0.010}^{+0.011}$   \\
$kT_{\rm comp}$ \dotfill                       &   keV                     & \ldots                      & $129_{-5}^{+4}$         & $16_{-13}^{+144}$           &  \ldots                      \\
$\tau$ \dotfill                                &                           & \ldots                      & $<0.41$                 & $< 9.1$                     &  \ldots                      \\ 
$\Gamma$ \dotfill                                       &                           & \ldots                      & \ldots                  & \ldots                      &  0.5$^\star$                 \\
$E_{\rm cut}$  \dotfill                                 &   keV                     & \ldots                      & \ldots                  & \ldots                      &  8.1$^\star$                 \\
Norm$_{\rm comp}$ \dotfill                     &  $10^{-5}$                & \ldots                      & $1.0_{-0.03}^{+6.9}$    & $3.9_{-3.5}^{+11}$          &  $6.71 \pm 0.48$             \\
$C_{\rm MOS1}$ \dotfill                                 &                           & $0.93 \pm 0.01$             & $0.93 \pm 0.01$         & $0.93 \pm 0.01$             &  $0.93 \pm 0.01$             \\ 
$C_{\rm MOS2}$  \dotfill                                &                           & $0.95 \pm 0.01$             & $0.94 \pm 0.01$         & $0.95 \pm 0.01$             &  $0.95 \pm 0.01$             \\
$C_{\rm FPMA}$ \dotfill                                 &                           & $1.08 \pm 0.02$             & $1.19 \pm 0.02$         & $1.12 \pm 0.03$             &  $1.12 \pm 0.03$             \\
$C_{\rm FPMB}$ \dotfill                                 &                           & $1.07 \pm 0.02$             & $1.18 \pm 0.02$         & $1.11 \pm 0.03$             &  $1.11 \pm 0.03$             \\ 
$\chi^2/\nu \; (\nu)$  \dotfill                            &                           & 1.23 (1714)                 & 1.21 (1712)             & 0.99 (1711)                 &  0.99 (1713)                 \\
Null $P$  \dotfill                                      &                           & $5.4 \times 10^{-11}$       & $6.7 \times 10^{-9}$    & 0.62                        &  0.56                        \\
\hline
\hline
\end{tabular}
\end{center}
 Note---The Galactic column density $N_{\rm H, Gal} = 5.7 \times
10^{20}$~cm$^{-2}$ has been applied to all of the above fits.  All quoted
errors are at the 90\% confidence level.\\
$^\star$Indicates parameter was fixed to the listed value (see $\S$3.1 for details).
\end{table*}

\section{Analysis and Results} \label{sec:results}

\subsection{Simultaneous {\itshape NuSTAR} plus {\itshape XMM-Newton} Spectral Fitting}

We began by fitting our nearly simultaneous \nustar\ and \xmm\ observations
using {\ttfamily XSPEC}~v.~12.9.1 (Arnaud~1996).  All fits were performed using
the \xmm\ pn, MOS1, and MOS2, and \nustar\ FPMA and FPMB data as described in
$\S$2.    To account for cross-calibration uncertainties and potential small
flux variations (\nustar\ and \xmm\ data were not taken exactly
simultaneously), we made use of a free {\ttfamily CONSTANT} model, which we
chose to be fixed at a value of 1.0 for the \xmm\ pn data.  Source and
background counts were binned as described in $\S$2, and source spectra were
fit by minimizing the $\chi^2$ statistic, after background counts were
subtracted.  For all \xmm\ data, the background is estimated to be a factor of
$>$10 times lower than the source.  For \nustar, the background remains a
factor of 5 times lower that the source for $E < 20$~keV, and is comparable to
the source level at $E =$~20--30~keV, where background emission lines are known
to be present.

%
%%%%%%%%%%%%%%%%%%%%%%%%%%%%%%%%%%%%%%%%%%%%%%%%%%%%%%%%%%%%%%%%%%%%%%%%%%%%%%%%%%
% Figure 4
%%%%%%%%%%%%%%%%%%%%%%%%%%%%%%%%%%%%%%%%%%%%%%%%%%%%%%%%%%%%%%%%%%%%%%%%%%%%%%%%%%
%
\begin{figure*}
\figurenum{4}
\centerline{
\includegraphics[width=9.0cm]{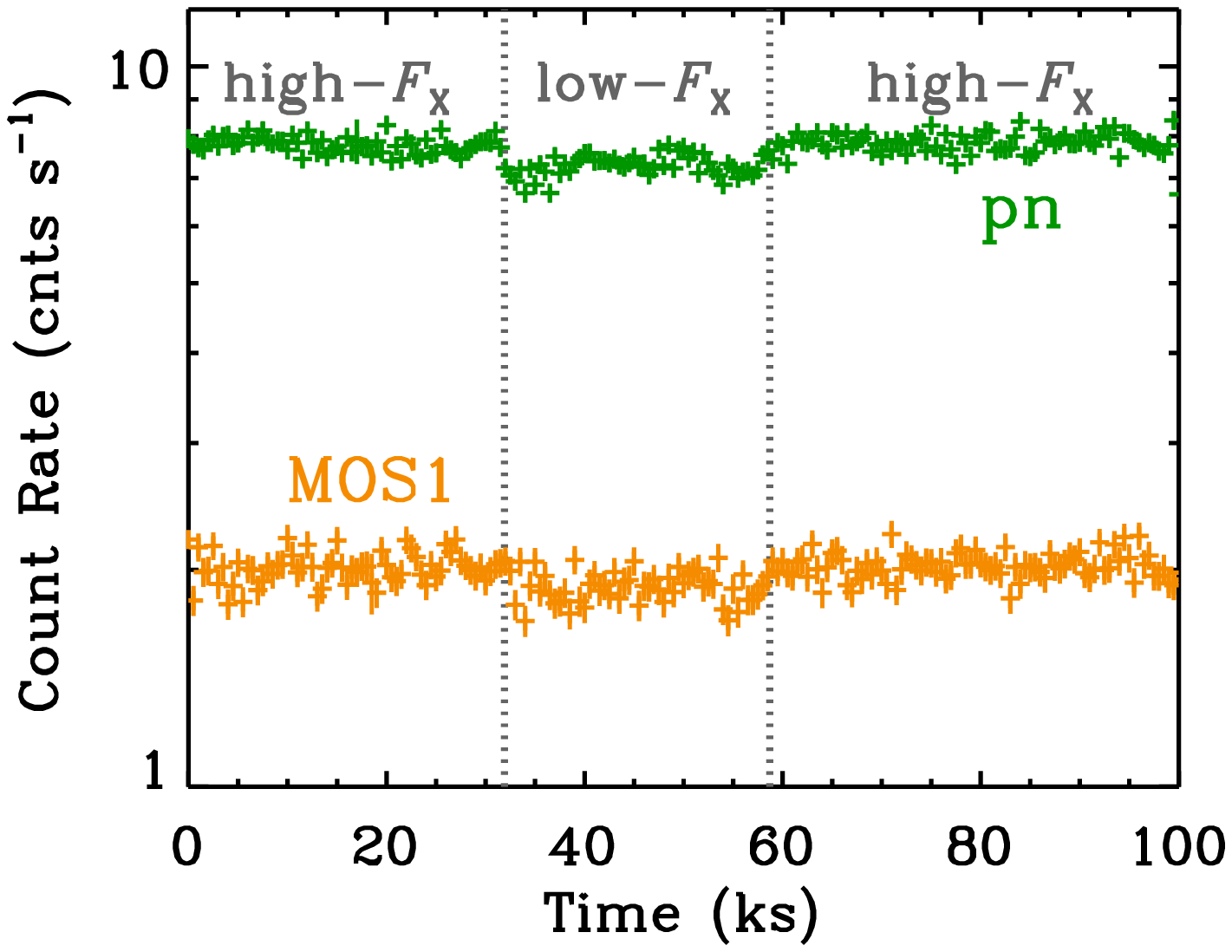}
\hfill
\includegraphics[width=9.0cm]{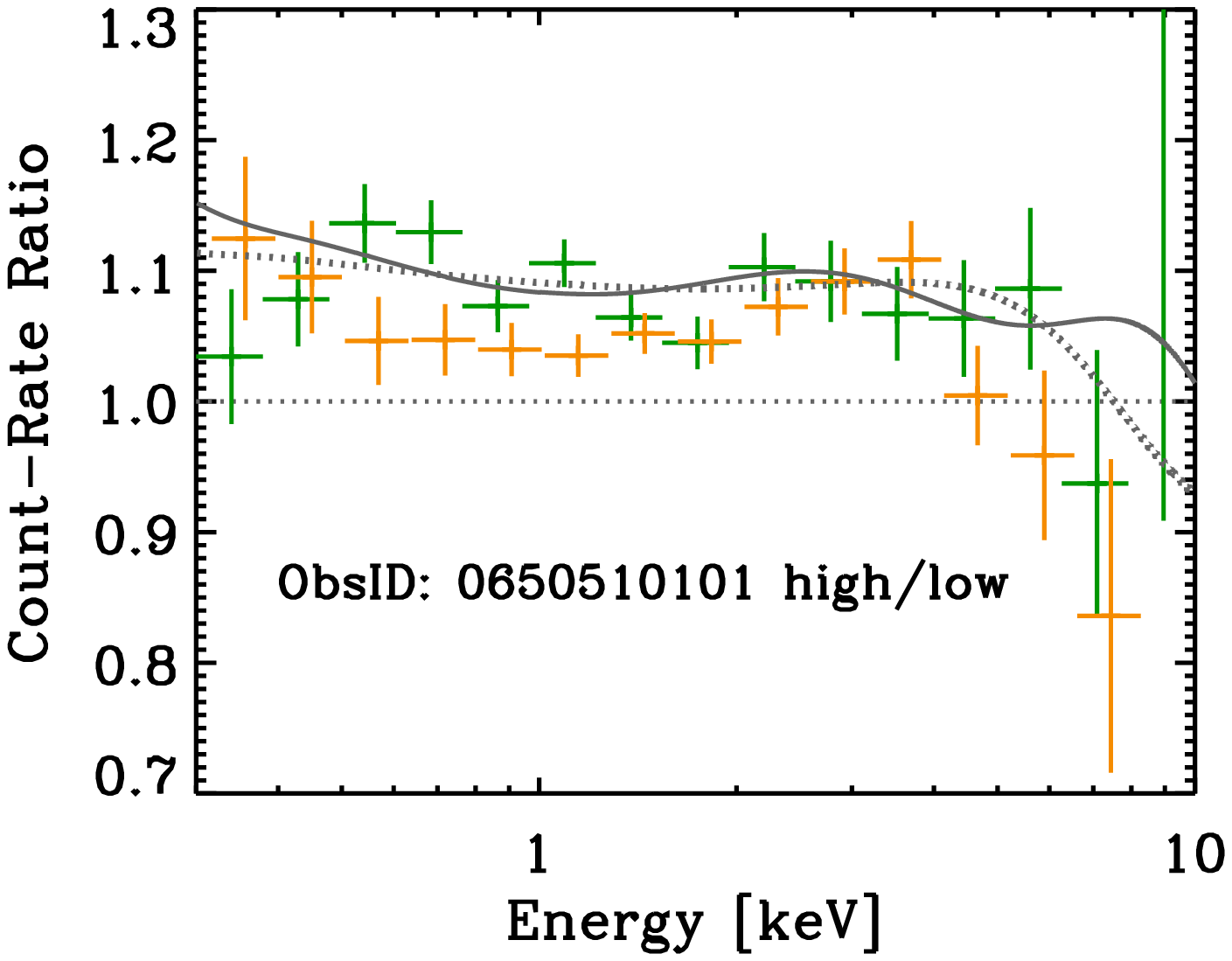}
}
\caption{
%%%
({\it a\/}) Light curve for \xmm\ ObsID: 060510101 for pn ({\it green\/})
and MOS1 ({\it orange\/}).  The light curve for MOS2 is very similar to that of
MOS1 shown here.  A $\sim$6--7\% ``dip'' in the light curves is observed
between 31.9 and 58.9~ks, as denoted by the vertical lines with annotated
periods of high and low flux.
%%%
({\it b\/}) Energy-dependent count-rate ratio between the high/low flux
periods for pn ({\it green\/}) and combined MOS ({\it orange\/}) detectors.
The differences in count-rate ratio are most obvious in the $\approx$0.3--4~keV
range in energy, suggesting variations in the accretion disk itself are
primarily responsible for the changes in flux, rather than in the
Comptonization or absorption.  Predicted ratios for the BH and NS baseline
models are shown with the solid and dotted curves, respectively.
%%%
}
\end{figure*}
%%%%%%%%%%%%%%%%%%%%%%%%%%%%%%%%%%%%%%%%%%%%%%%%%%%%%%%%%%%%%%%%%%%%%%%%%%%%%%%%%%

We chose to fit the spectra using both ``BH'' and ``NS'' models.  For the BH
models, we followed M11, who found two super-Eddington ({\ttfamily TBABS
$\times$ (DISKPBB + COMPTT)} and {\ttfamily TBABS $\times$ DISKPBB}) and one
sub-Eddington ({\ttfamily TBABS $\times$ (DISKBB + COMPTT)}) models provided
good fits to the \xmm\ data available at the time.  Phenomenologically, the
{\ttfamily DISKPBB} component serves as a model of the visible contribution of
an advection dominated accretion disk, expected for accretion disks near the
Eddington limit (e.g., Mineshige \etal\ 1994; Kubota \& Makishima~2004).  The
model consists of an accretion disk with a radial temperature gradient,
following $T(r) \propto r^{-p}$, where $p = 0.75$ is the case of a standard
thin disk and $p =$~0.5--0.75 would be the case for an advection-dominant disk.
The {\ttfamily DISKBB} model serves as a standard geometrically thin accretion
disk (e.g., Kubota \etal\ 1998). The {\ttfamily COMPTT} model calculates the
spectrum of seed photons as seen through a Comptonizing screen, which are
assumed to be generated by some portion of the accretion disk. For all relevant
fits, we establish a connection between the accretion disk and Comptonization
components of the models by linking the seed-photon temperature of the
{\ttfamily COMPTT} model to the temperature of the inner disk of either the
{\ttfamily DISKPBB} or {\ttfamily DISKBB} models.

For the NS case, we fit the spectra to the super-Eddington model from W18
({\ttfamily TBABS $\times$ (DISKPBB + DISKBB + CUTOFF-PL)}).  Here, the
accretion disk is thought to be composed of an inner advection dominant portion
close to the NS ({\ttfamily DISKPBB}) and a standard outer portion ({\ttfamily
DISKBB}).  Following W18, we required that the {\ttfamily DISKBB} component
yield an inner temperature below 1~keV to prevent swapping of temperatures with
the hotter {\ttfamily DISKPBB} component.  However, we found that the
{\ttfamily DISKBB} component was not required in the case of M33 X-8;
hereafter, we made use of the {\ttfamily TBABS $\times$ (DISKPBB + CUTOFF-PL)}
for the NS case.  The {\ttfamily CUTOFF-PL} component models Comptonization
that is thought to arise from an accretion column falling onto the poles,
within the magnetospheric radius.  All known ULX pulsars have accretion columns
that show broadly consistent spectral shapes (e.g., Brightman \etal\ 2016;
W18).  In our fitting, we followed the procedure of W18 and fixed the
{\ttfamily CUTOFF-PL} shape to be the average of that measured from the three
ULX pulsars known at the time (M82 X-2, NGC 7793 P13, and NGC~5907 ULX1).  This
model includes a photon-index of $\Gamma = 0.5$ and cut-off energy of $E_{\rm
cut} = 8.1$~keV.  In our procedure, we vary only the normalization of this
component.

In Figure~3, we show the \xmm\ pn and MOS, and \nustar\ FPMA and FPMB spectrum of M33
X-8, along with residuals to fits using the four models.  In Table~2, we list
the best fit model parameters and goodness-of-fit values.  Thanks to the
\nustar\ constraints, we can uniquely rule out both the pure advection
dominated disk ({\ttfamily DISKPBB}) and sub-Eddington ({\ttfamily
DISKBB+COMPTT}) BH models on the grounds of their poor fits (null-hypothesis
probabilities of $P < 6.7 \times 10^{-9}$), particularly for the $E >$~10~keV
data.  Instead, our data find the {\ttfamily DISKPBB+COMPTT} BH model and the
{\ttfamily DISKPBB + CUTOFF-PL} NS model acceptable with null-hypothesis
probabilities of 0.62 and 0.56, respectively (see Table~2).  

For completeness, we also performed fits to additional basic spectral models
that are often used to describe ULX spectra (e.g., {\ttfamily POWERLAW},
{\ttfamily BKNPOWERLAW}, {\ttfamily DISKBB + POWERLAW}, and {\ttfamily
DISKBB+CUTOFFPL}).  Fit parameters based on these models are provided in
Appendix~A.  We find that most of these models were either poor fits to the data or
unlikely on physical grounds.  For example, the {\ttfamily DISKBB + POWERLAW}
and {\ttfamily DISKBB + CUTOFFPL} only provide reasonable fits when the
power-law components are significantly more luminous than the disk component at
low energies ($\simlt$2~keV).  Such a solution is possible for sources where the seed photons arise from the Comtonizing source itself (e.g., an accretion column onto a pulsar). However, such sources are lower luminosity and exhibit obvious pulsations, which we do not find for X-8 (see $\S$3.4). 

In the next section, we adopt the {\ttfamily DISKPBB+COMPTT} BH model and
{\ttfamily DISKPBB + CUTOFF-PL} NS model when fitting archival \xmm\ data;
hereafter, we refer to these models as ``baseline'' BH and NS models,
respectively.  For the BH baseline model, we found than the {\ttfamily COMPTT}
temperature  ($kT_{\rm comp}$) and optical depth ($\tau$) were highly
correlated and poorly constrained, even for our broad band fits.  Our model
suggests that a more optically-thick Comptonization component is slightly
preferred.  Therefore, when fitting to archival \xmm\ data, we chose to fix
$\tau = 0.75$, near the best-fit value from the broad band fitting, and fit for
$kT_{\rm comp}$.

\subsection{Spectral Fits to Individual {\itshape XMM-Newton} Archival
Observations}

As discussed in $\S$2.2, a full list of the \xmm\ archival data is provided in
Table~1, with the timeline of X-8 source fluxes mapped in Figure~2.  The fluxes
reported are based on spectral fits to the pn data using our BH baseline model
for each observation, with 90\% uncertainties on the fluxes displayed.  We note
that the cross-calibration fluxes between instruments (pn, MOS, and \nustar\
detectors) is at the $\approx$7--10\% level, suggesting that absolute
calibration uncertainty is generally larger that the errors on the fluxes
quoted here (e.g., Read \etal\ 2014; Madsen \etal\ 2017); however, the relative flux errors should be insensitive to this.
The
0.3--10~keV flux has undergone little variation over the last 17 years of \xmm\
observations, with the observational mean and standard deviation of $F_{\rm X}
= (1.67 \pm 0.11) \times 10^{-11}$~\flux\ ($L_{\rm X} = [1.38 \pm 0.09] \times
10^{39}$~\lum).  Such a small scatter ($\approx$6\%) in flux differs from the
$\approx$30\% reported by La~Parola \etal\ (2015), which was based on fluxes
from \xmm, \chandra, \suzaku, \sax, and \swift.  Their reported scatter is
dominated by the previously reported \xmm\ fluxes by M11, which are exclusively
from MOS-camera data.  The most extreme fluxes presented in the M11 study are
based on \xmm\ observations that were reported to have issues (e.g., slew
failure, high levels of background radiation, and telemetry glitches).  For
example, M11 report that X-8 had an all time low MOS1 flux of $F_{\rm X}
\approx 1.1 \times 10^{-11}$~\flux\ during the 2003 Jan 22 observation
(ObsID:0141980501); however, this observation was noted to suffer from a slew
failure and resulted in large differences in calibration between cameras ---
the pn flux for X-8 during this observation is $F_{\rm X} \approx 1.8 \times
10^{-11}$~\flux.  Using only the 14 \swift\ fluxes reported in Table~1 of
La~Parola \etal\ (2015), we obtain a scatter of only $\approx$7.7\%, comparable
to that reported here. Past monitoring from \rosat\ has suggested that X-8 may exhibit a modulated period of $\approx$106~days with an amplitude of
$\approx$20\% (Dubus \etal\ 1997).  The variability over the last
17 years is consistent with this amplitude.

Despite the small variations in flux, some spectral variations exist in the
historical \xmm\ observations of X-8, and these variations are broadly
correlated with source flux.  Using our BH and NS baseline models, we fit the data for
each of the 17 \xmm\ observations listed in Table~1 and attempted to isolate
parameters of the model.  The majority of our observations have relatively
short exposures ($\simlt$10~ks), and therefore the parameters of our models
are only poorly constrained.  Nonetheless, both BH and NS models provide a
good-to-sufficient fit to the data (null-hypothesis
probabilities of ~0.02--0.75) for all ObsIDs, with
the exception of ObsID: 0650510101 (null-hypothesis
probability of $10^{-4}$), which constitutes one
of the two longest observations in the archive (see Williams \etal\ 2015).

During ObsID: 0650510101, X-8 was reported by Sutton \etal\ (2013) to have a
fractional variability of $F_{\rm var} = 3.1 \pm 0.3$\% (calculated following
Vaughan \etal\ 2003), which is $\approx$4 times larger than that found in the
comparably long observation ObsID: 0650510201 ($F_{\rm var} = 0.7 \pm 0.5$\%).
In Figure~4$a$ we show the light curve for ObsID: 0650510101.  It is clear that
the large fractional variability is dominated by a $\approx$27~ks dip in the
flux of X-8 that takes place $\approx$32~ks after the start of the observation.
We investigated how the spectral properties of X-8 changed across the
observation by extracting a ``low-flux'' spectrum at interval 31.9--58.9~ks and
a ``high-flux'' spectrum, combining intervals 0--31.9~ks plus 58.9-100~ks (see
Fig.~4$a$ annotations).  The high flux 0.3--10~keV count rate is observed to be
7.2\% and 6.4\% higher than the low-flux count rate for the pn and mean MOS1+MOS2
detectors, respectively.

%%%%%%%%%%%%%%%%%%%%%%%%%%%%%%%%%%%%%%%%%%%%%%%%%%%%%%%%%%%%%%%%%%%%%%%%%%%%%%%%%%
% Table 3
%%%%%%%%%%%%%%%%%%%%%%%%%%%%%%%%%%%%%%%%%%%%%%%%%%%%%%%%%%%%%%%%%%%%%%%%%%%%%%%%%%
\begin{table*}
\begin{center}
\caption{Best Fit Parameters for Grouped Archival \xmm\ Observations.}
\begin{tabular}{lccccccc}
\hline\hline
\multicolumn{2}{c}{} & \multicolumn{2}{c}{ObsID 0650510101 Split}  & \multicolumn{4}{c}{0.3--10~keV Flux Range (10$^{-11}$~\flux)}  \\ 
\multicolumn{2}{c}{} &  \multicolumn{2}{c}{\rule{1.5in}{0.01in}} & \multicolumn{4}{c}{\rule{3.0in}{0.01in}}\\ %
\multicolumn{1}{c}{\sc Parameter} & {\sc Unit} & (low-$F_{\rm X}$) & (high-$F_{\rm X}$) & ($<$1.5) & (1.5--1.6) & (1.6--1.7) & ($>$1.7) \\ 
\hline
Bin \dotfill                                   &  & \ldots & \ldots & 1 & 2 & 3 & 4  \\
$N_{\rm Bin}^\dagger$ \dotfill                         &  & 1 & 1 & 1 & 4 & 6 & 5 \\
\hline
\multicolumn{8}{c}{BH Baseline Model} \\
\hline
$N_{\rm H}$ \dotfill                           &   10$^{22}$ cm$^{-2}$    & 0.157 $\pm$ 0.000 & 0.157$_{-0.016}^{+0.030}$ & 0.161$_{-0.014}^{+0.015}$ & 0.147$_{-0.007}^{+0.008}$ & 0.166$_{-0.014}^{+0.012}$ & 0.164$_{-0.013}^{+0.017}$ \\
$kT_{\rm in}$ \dotfill                         &   keV                     & 0.99$_{-0.32}^{+0.25}$ & 0.72$_{-0.08}^{+0.20}$ & 1.01$_{-0.28}^{+0.55}$ & 0.84$_{-0.05}^{+0.09}$ & 0.82$_{-0.08}^{+0.13}$ & 0.94$_{-0.15}^{+0.60}$ \\
$p$ \dotfill                                   &                           & 0.556 $\pm$ 0.006 & 0.547$_{-0.033}^{+0.028}$ & 0.517$_{-0.017}^{+0.016}$ & 0.556$_{-0.056}^{+0.012}$ & 0.529$_{-0.029}^{+0.020}$ & 0.537$_{-0.037}^{+0.018}$ \\
$(R_{\rm in}/D_{10})^2 \cos \theta$ \dotfill   &                           & 0.33$_{-0.20}^{+1.02}$ & 1.04$_{-0.53}^{+0.37}$ & 0.18$_{-0.14}^{+0.42}$ & 0.55$_{-0.14}^{+0.11}$ & 0.50$_{-0.17}^{+0.15}$ & 0.34$_{-0.29}^{+0.22}$ \\
$kT_{\rm Comp}$ \dotfill                       &   keV                     & $<$28.7 & 18.2$_{-1.8}^{+5.5}$ & $<$62.8 & 15.3$_{-0.9}^{+1.0}$ & 14.0$_{-1.9}^{+1.7}$ & $<$3.1 \\
Norm$_{\rm Comp}$ \dotfill                     &   10$^{-4}$               & 0.30$_{-0.28}^{+0.74}$ & 1.12$_{-0.66}^{+0.39}$ & 0.18$_{-0.14}^{+0.51}$ & 0.85$_{-0.27}^{+0.18}$ & 1.08$_{-0.35}^{+0.25}$ & 0.89$_{-0.89}^{+0.34}$ \\
$F_{\rm X}^{\rm disk}/F_{\rm X}$               &                           & 0.75 & 0.55 & 0.67 & 0.64 & 0.60 & 0.67 \\
$\chi^2/\nu (\nu)$                             &                           & 1.00 (984)  & 1.09 (1705)  & 1.03 (829)  & 1.03 (4955)  & 1.05 (3335)  & 1.10 (2577)  \\
Null $P$                                       &                           & 0.4820 & 0.0043 & 0.2413 & 0.0490 & 0.0233 & 0.0004 \\
\hline
\multicolumn{8}{c}{NS Baseline Model} \\
\hline
$N_{\rm H}$ \dotfill                           &   10$^{22}$ cm$^{-2}$    & 0.148 $\pm$ 0.000 & 0.148$_{-0.009}^{+0.008}$ & 0.167$_{-0.016}^{+0.006}$ & 0.142 $\pm$ 0.004 & 0.152$_{-0.007}^{+0.006}$ & 0.160$_{-0.008}^{+0.006}$ \\
$kT_{\rm in}$ \dotfill                         &   keV                     & 1.11$_{-0.06}^{+0.07}$ & 1.16$_{-0.06}^{+0.07}$ & 1.24$_{-0.14}^{+0.16}$ & 1.24 $\pm$ 0.03 & 1.36$_{-0.08}^{+0.09}$ & 1.53$_{-0.14}^{+0.12}$ \\
$p$ \dotfill                                   &                           & 0.561 $\pm$ 0.005 & 0.558$_{-0.010}^{+0.012}$ & 0.502$_{-0.002}^{+0.016}$ & 0.559 $\pm$ 0.005 & 0.545$_{-0.007}^{+0.008}$ & 0.540$_{-0.006}^{+0.005}$ \\
$(R_{\rm in}/D_{10})^2 \cos \theta$ \dotfill   &                           & 0.21$_{-0.05}^{+0.06}$ & 0.19$_{-0.05}^{+0.06}$ & 0.06$_{-0.01}^{+0.05}$ & 0.13$_{-0.01}^{+0.02}$ & 0.09$_{-0.02}^{+0.03}$ & 0.06$_{-0.01}^{+0.03}$ \\
Norm \dotfill                                  &   10$^{-4}$               & 2.13$_{-0.50}^{+0.44}$ & 1.90$_{-0.37}^{+0.34}$ & 2.73$_{-0.47}^{+0.32}$ & 1.15$_{-0.13}^{+0.12}$ & 0.87$_{-0.44}^{+0.37}$ & 0.51$_{-0.51}^{+0.68}$ \\
$\chi^2/\nu (\nu)$                             &                           & 1.00 (985)  & 1.09 (1706)  & 1.04 (830)  & 1.03 (4956)  & 1.05 (3336)  & 1.10 (2578)  \\
Null $P$                                       &                           & 0.4837 & 0.0039 & 0.2331 & 0.0446 & 0.0202 & 0.0004 \\
\hline
\hline
\end{tabular}
\end{center}
$^\dagger$ Number of \xmm\ ObsIDs in a given bin.\\
Note---All quoted errors are at the 90\% confidence level.\\
\end{table*}

%
%%%%%%%%%%%%%%%%%%%%%%%%%%%%%%%%%%%%%%%%%%%%%%%%%%%%%%%%%%%%%%%%%%%%%%%%%%%%%%%%%%
% Figure 5
%%%%%%%%%%%%%%%%%%%%%%%%%%%%%%%%%%%%%%%%%%%%%%%%%%%%%%%%%%%%%%%%%%%%%%%%%%%%%%%%%%
%
\begin{figure*}
\figurenum{5}
\centerline{
\includegraphics[width=16.0cm]{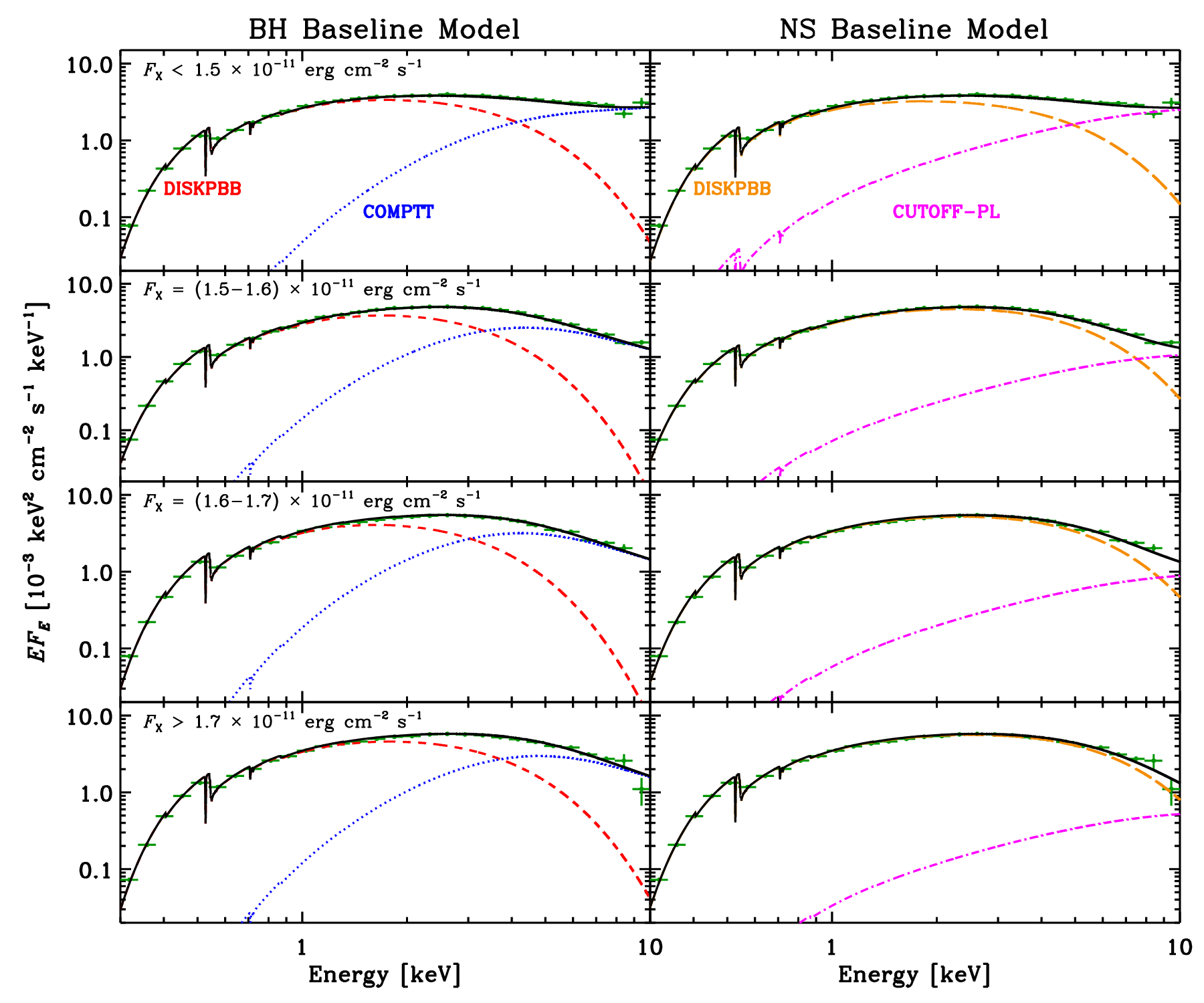}
}
\caption{
%%%
0.3--10~keV \xmm\ $EF_E$ spectra of M33 X-8, grouped by flux range for both our
BH ({\it left column\/}) and NS ({\it right column\/}) baseline model.
Best-fit models ({\it black curves\/}) were constructed using the joint
spectral fitting technique described in $\S$3.3 and fit parameters are provided
in Table~3.  Green data points represent energy-binned average values and are
plotted here for ease of viewing (these were not used in our spectral fitting).
For the BH baseline models, the {\ttfamily DISKPBB} ({\it red dashed curves\/})
and {\ttfamily COMPTT} ({\it blue dotted curves\/}) contributions are shown,
and for the NS baseline models, we display the {\ttfamily DISKPBB} ({\it orange
long-dashed curves\/}) and {\ttfamily CUTOFF-PL} ({\it magenta dot-dashed
curves\/}) contributions. 
%%%
}
\end{figure*}
%%%%%%%%%%%%%%%%%%%%%%%%%%%%%%%%%%%%%%%%%%%%%%%%%%%%%%%%%%%%%%%%%%%%%%%%%%%%%%%%%%

In Figure~4$b$, we show the high-to-low count-rate ratio as a function of
energy.  This ratio was computed by binning the spectra to evenly-divided $\log
E$ bins.  From these data, it appears that the predominant changes in the
spectrum took place across the $\approx$0.3--4~keV energy range with the
$\simgt$4~keV data appearing to stay roughly constant between the two
observations.  This suggests that the changes in count rate were most likely
associated with the accretion disk itself and not related to changes in
absorption and Comptonization.  To test this, we first performed fits to the
high-flux spectrum using both our BH and NS baseline models.  Fixing the
absorption ({\ttfamily TBABS}) and Comptonization (either {\ttfamily COMPTT} or
{\ttfamily CUTOFF-PL}) components, we then fit the low-flux spectrum,
varying only the disk-component temperature, $p$ value, and normalization.
Figure~4$b$ shows the predicted count-rate ratio as a function of energy for
both the BH ({\it solid curve\/}) and NS ({\it dotted curve\/}) baseline
models.  Both BH and NS baseline models provided statistically acceptable fits
to the data (see Table~3), suggesting indeed that the dip in flux can be described well
by changes associated with the accretion disk.  For the BH baseline model, the
disk appears to be cooler and truncated (i.e., larger $R_{\rm in}$) in the
low-flux case.  The NS baseline model favors a slight cooling of the disk with
the radial temperature profile becoming somewhat steeper (i.e., an increase in
$p$).  

%Exceptions to this are the observations conducted in 2010 (see Williams \etal\
%2015), which are $\simgt$100~ks, and the new observations from 2017 reported
%here (see Table 1 for exposure times used in each observation).  
\subsection{Flux-Binned Spectral Fits to {\itshape XMM-Newton} Archival
Observations}

For the purpose of obtaining the best possible constraints on the spectral
variations, we grouped the \xmm\ X-8 spectra according to their
\hbox{0.3--10~keV} flux and performed joint spectral fitting of the data within
a given bin.  The observations were arbitrarily divided into four flux bins of
$F_{\rm X}$/($10^{-11}$~\flux) $< 1.5$, 1.5--1.6, 1.6--1.7, and $> 1.7$, which
we hereafter refer to as bins~1, 2, 3, and 4, respectively.  The bin
assignments for the individual ObsIDs are listed in Table~1 and graphically
illustrated in Figure~2 with the plotted circles colored according to their
designated flux bin.  Bins~1, 2, 3, and 4 respectively contain 1, 4, 6, and 5
unique \xmm\ ObsIDs.  In this exercise, we exclude ObsID: 0650510101, which was
presented in the previous section, to provide unique results.

In our joint spectral fitting procedure, all data in a given bin were fit to
our BH and NS baseline models to common (i.e., linked) model parameters, with
the exception of a multiplicative {\ttfamily constant} model that was allowed
to vary for each ObsID.   In Table~3, we list the resulting best-fit parameters
and goodness of fit for each of the four bins.  
For illustrative purposes, in Figure~5 we
show binned $E F_E$ spectra for the four bins and the best-fit models, with
model components separated.  The stacked spectra shown in Figure~5 ({\it green
points\/}) are average values with 1$\sigma$ errors on the mean of all 
data points within evenly-divided $\log E$ bins.  Note that these stacked
values were not used in our spectral fitting procedure and are simply shown
here for illustrative purposes.

Both the BH and NS baseline models provide reasonable and nearly equivalent quality fits to
the data; however, there is some tension in the fit quality for the data in
the highest-flux bin (Bin 4), where the null-hypothesis
probability is $0.0004$ for both models. Examination of the residuals to these fits suggests that there is likely some correlation between flux and the spectral shape within a given flux bin. 
As noted in $\S$3.2, all ObsIDs, except for ObsID 0650510101, can be fit well by these
models on an individual basis.  Thus, changes in the physical properties of the
sources are likely responsible for the relatively low values of null-hypothesis
probability here.   

The best-fit parameters from our fits to the binned archival \xmm\ data show only small changes over the range of \xray\ fluxes covered. Nonetheless, some basic trends are apparent.  In general, one can observe from Figure~5 that as $F_{\rm X}$ increases, the curvature of the higher-energy (i.e., $E \approx$~2--10~keV) portion of the spectrum covered by \xmm\  increases, going from a relatively flat spectrum at $F_{\rm X}$/($10^{-11}$~\flux) $< 1.5$ to one that shows a clear $E \approx$~1--3~keV turnover by $F_{\rm X}$/($10^{-11}$~\flux) $> 1.7$.  The shape of the low-energy ($\simlt$1~keV) spectrum, which we associate with the accretion disk, undergoes more subtle changes.  Figure~6 graphically displays best-fit parameter values for the accretion disk versus $F_{\rm X}$ for both the BH and NS cases. Although subject to large uncertainties and some degeneracies between model parameters, we can use trends in the data here to make tentative interpretations of the nature of X-8.  
%\added{
For guidance, in Figure 6, we show least squares regression lines fit to all the displayed data. For the BH baseline model, 
%\deleted{the flux of the accretion disk component seems to decrease with increasing $F_{\rm X}$, necessitating an increase in the amount of flux contributed by the Componization component. Thus,} 
the flux-dependent variations in the spectrum of X-8
%\replaced{can be interpreted as (1) the Comptonization component becoming more luminous with increasing $F_{\rm X}$ (a proxy for accretion rate); and (2) the visible disk component becoming truncated and thus cooler as $F_{\rm X}$ increases.  These results are consistent with those}
are consistent with the interpretation presented by M11, who proposed that as the accretion rate increases, winds from the disk become driven more effectively at larger radii due to the increased radiation pressure.  The wind itself would provide a source of Comptonization This prediction leads to lower $kT_{in}$, larger inner radii, and a declining $F_{disk}/F_{\rm X}$ fraction with increasing $F_{\rm X}$. While the data do not show statistically significant correlations with $F_{\rm X}$, we do find that the preferred slopes from our least squares regression fits are consistent with this picture. 

The NS baseline model case provides a somewhat different picture. Here X-8 can be interpreted as becoming more disk dominant with increasing flux, with the temperature (and possibly the size) of the accretion disk increasing.  In comparison to other ULXs with broad band constraints from W18a, such behavior is similar to that observed for the ULX Holmberg IX X-1, which shows clear increase in curvature with increasing flux, most likely due to changes in the accretion disk (Luangtip \etal\ 2016; Walton \etal\ 2017).  However, the known pulsar ULX, NGC 5907 ULX1 shows an increase in the high-energy component with increasing flux associated with the pulsar
accretion column (W18a; F{\"u}rst \etal\ 2017).  W18a speculate that the contrasting behavior of Holmberg IX X-1 with that of NGC~5907 ULX1 could be an indication of differing compact-object types (i.e., BH vs.~NS) or viewing angles.

%
%%%%%%%%%%%%%%%%%%%%%%%%%%%%%%%%%%%%%%%%%%%%%%%%%%%%%%%%%%%%%%%%%%%%%%%%%%%%%%%%%%
%Figure 6
%%%%%%%%%%%%%%%%%%%%%%%%%%%%%%%%%%%%%%%%%%%%%%%%%%%%%%%%%%%%%%%%%%%%%%%%%%%%%%%%%%
%
\begin{figure} 
\figurenum{6} 
\centerline{
\includegraphics[width=9.0cm]{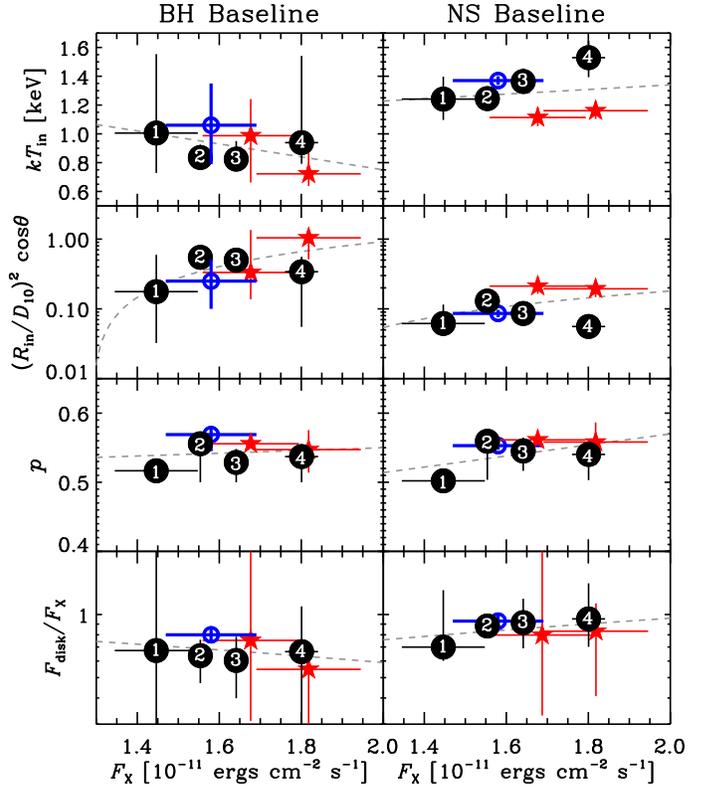} } 
\caption{
%%%
Accretion disk parameters versus 0.3--10~keV flux ($F_{\rm X}$) for the BH
({\it left column\/}) and NS ({\it right column\/}) baseline models.  These
plots include, from top to bottom, the inner accretion disk temperature,
radius, temperature profile index $p$, and the fraction of the total flux
attributed to the accretion disk.  
Fit results are shown for our nearly simultaneous \nustar\ plus \xmm\ exposure ({\it open blue circle\/}), the high/low flux intervals of ObsID 0650510101 ({\it red stars}), and $F_{\rm X}$-divided bins ({\it black circles with annotated bin number\/}).
%\added{
Least squares regression lines ({\it dotted gray lines\/}) fit to all data points in each panel are included as indicators of potential trends with $F_{\rm X}$.
%Albeit with large uncertainties, our results are
%consistent with the accretion disk portion of the spectrum appearing cooler
%with increasing flux ($a$), possibly due to an increasingly prominent
%optically-thick outflow that obscures the inner, hotter portions of the
%accretion disk ($b$).
%%%
} 
\end{figure}
%%%%%%%%%%%%%%%%%%%%%%%%%%%%%%%%%%%%%%%%%%%%%%%%%%%%%%%%%%%%%%%%%%%%%%%%%%%%%%%%%%

\subsection{Search for Pulsations}

While our spectral analyses provide new constraints on the nature of accretion
in M33 X-8, we cannot draw firm conclusions about the nature of the compact
object itself.  Given the discovery of pulsations in other ULXs, we searched
the 3--20~keV light curves of the two {\it NuSTAR} epochs that contain X-8 (ObsIDs 50310002001 and 50310002003) to test whether the hard component is pulsating.  

%{\it XMM-Newton} light curves are extracted 
%using {\it XMM} {\tt SAS} version 16.1.0 
%from 0.2--12~keV and 3-8 keV EPIC pn data using time bins of 0.08s and 100s. Circular regions of radii $\sim$50$\arcsec$ and $\sim$100$\arcsec$ are used for the source and clean background areas, respectively. We searched for periodicities \color{red} (in what range?) \color{black} between 0.15s and $\sim$22 ks in each of the {\it XMM-Newton} observations with frequency resolutions of 0.01, 0.001 and 0.0001 Hz. 

%We extract {\it NuSTAR} light curves from the 3--20~keV energy range using time bins of \color{red} Xs and Ys\color{black}. % using {\tt nuproducts} in {\tt HEASoft} version~6.22. 
%\color{red} NuSTAR light curve extraction details???\color{black}

%{\it NuSTAR}'s time resolution is good to $\sim$2~ms rms after being corrected for thermal drift of the on-board clock (Fornasini \etal\ 2017), and the absolute accuracy of the time resolution can be better than 3~ms (Mori \etal\ 2014; Madsen \etal\ 2015).

All photon arrival times were converted to barycentric dynamical time. Following the methodology of Yang \etal\ (2017) in search of periodic pulsations, we constructed Lomb-Scargle periodograms 
%over the frequency range \color{red} {\bf X-Y} \color{black} 
for the whole observing time and searched for spin periods in the range 0.15--200476~s.
%, with a logarithmic scaling of {\bf Z}
No significant pulsations were detected in the second {\it NuSTAR} epoch. However, a periodic signal was independently detected in both telescopes FPMA ($769.42 \pm 4.87$~s) and FPMB ($724.82 \pm 5.22$~s) during the first {\it NuSTAR} epoch (ObsID 50310002001), but the periods are statistically inconsistent. This should not be the case if a true periodic signal was emanating from the source, given that the source was observed simultaneously by these two telescopes. 

% Still need to condense this paragraph %
The discrepancy between these candidate period values is not explained by any typical observing window function effect, but
%not likely to be an effect of the observing window function, as there are no corresponding $\sim$700~s signals detected in the background regions of the observations. It is also possible that one of the detected periods is a real periodicity in the M33 X-8 light curve, while the other has been aliased by the window function (as discussed in VanderPlas~2018). However, neither aliasing by {\it NuSTAR}'s orbital period ($\sim$5280~s) nor by an SAA-related window function can cause the discrepancy. A perhaps more likely explanation is 
could be due to the presence of red noise that behaves as a broken power law in the light curve of M33 X-8. The break frequency could be broadly picked up as a single periodicity by the Lomb-Scargle algorithm, resulting in relatively similar but statistically distinct values being independently detected by the FPMA and FPMB telescopes.
%{\bf I reconstructed this from memory via emails with Tom -- he can restate this better
%and more correctly I suspect.}

Due to the inconsistency between period values and lack of any physically compelling explanation for the discrepancy, we cannot claim to have found a significant, coherent signal in the {\it NuSTAR} light curves searched in our study.

\section{Discussion and Conclusions} \label{sec:discussion}

The broad band spectra and light curves of M33 X-8, presented in $\S$3, provide
updated constraints on the nature of this source.  Thanks
to the nearly simultaneous \xmm\ and \nustar\ observational constraints, we can
now rule out the previously acceptable sub-Eddington ({\ttfamily DISKBB +
COMPTT}) and pure advection-dominated disk ({\ttfamily DISKPBB}) models for
X-8.  Instead, we find that the constraints are consistent with more recent
models of super-Eddington accretion onto either a BH or NS, albeit with
no clear preference for a BH or NS accretor.

While preparing this manuscript, a paper on M33 X-8 by Krivonos \etal\ (2018)
was published that made use of the \nustar\ data that we present here, in
combination with \swift\ archival data, to provide lower-energy constraints.
In general, we find basic agreement between our spectral constraints, and those
of Krivonos \etal\ (2018), with the exception that their data allows for an
acceptable fit using the {\ttfamily DISKBB + COMPTT}, which prompted them to
claim that M33 X-8 is most likely a BH XRB in a very high state.  We attribute
the difference in findings to their use of archival \swift\ data, which spans a
range of fluxes and provides weaker spectral constraints on the data.  Our use
of nearly simultaneous, high-S/N \xmm\ and \nustar\ data is therefore crucial
for ruling out this model.  Here, we prefer an interpretation in which M33 X-8
is in an accretion state in which the structure of the disk has been modified
by advection, as required by our broad band fits.  Such sources have been
purported to be likely BH XRBs due to their similarities with known Galactic BH
XRBs (Sutton \etal\ 2017);  however, we are unable to clearly distinguish
between a NS or BH nature of the accretor.

For the BH interpretation, the data presented here are consistent with past
phenomenological models of ULXs, in which an advection-dominated disk appears
cooler and truncated at relatively high fluxes, potentially due to the increase
in outflowing material from the inner portions of the disk.  
Here, the outflowing material
provides both
obscuration of the hot inner disk and Comptonization (see $\S\S$5.3 and 5.4 of
Kaaret \etal\ 2017 for a comprehensive description of this model).  In this
context, the increase in spectral curvature at $\simgt$3~keV with increasing
flux is expected to be due to the Comptonization component becoming cooler and
possibly more optically thick.  Other variable ULXs have been reported to show
this same type of behavior, when analyzing 0.3--10~keV spectra (most notably
with \xmm).  For example, a recent observation of a drop in the flux of ULX IC~342
X-1 showed a shift in the spectral turnover to higher energies as the
luminosity decreased by a factor of $\sim$2 (Shidatsu \etal\ 2017).  Similarly,
Ho~IX~X-1 has shown similar spectral changes over a broad range of fluxes.  For this source, the spectrum transitions smoothly from
a relatively flat
high-energy component at $\simgt$2~keV for low luminosities to 
a highly curved component at the highest luminosities of the source (Luangtip
\etal\ 2016).  

The above spectral behavior can also be interpreted as changes in the accretion
disk itself.  In the right column of Figure~5, we showed, using our NS baseline
model, that the 0.3--10~keV spectral variability of M33 X-8 could be modeled
well by an advection-dominated disk increasing in temperature with increasing
flux.  In this scenario, the increased importance of the accretion disk
relative to the high-energy Comptonization component is what leads to spectral
curvature at $\simgt$3~keV, instead of the Comptonization itself.  The
Comptonization component itself, which is modeled here as an accretion column
onto the NS, would have very minor contributions to the 0.3--10~keV spectra, and is thus
poorly constrained.  Furthermore, W18a showed that both Ho~IX~X-1 and IC~342
X-1, when analyzed using broad band (0.3--30~keV) spectral data, can also be
fit well using their NS-based two-component thermal model ({\ttfamily DISKPBB +
DISKBB}) plus Comptonization ({\ttfamily CUTOFF-PL}) model.  They too find that for
Ho~IX~X-1, for which broad band observations are available for different flux
states, the enhancement in spectral curvature with increasing flux can be
explained by a rise in the contribution of the {\ttfamily DISKPBB} component, with little change in
Comptonization.

It is important to point out here that the changes in the broad band spectra of
Ho~IX~X-1, as modeled by W18a, differ from those of the known pulsating ULX
NGC~5907 ULX1, which was shown to have strong variations in the hard
Comptonization component, associated with the accretion column of the NS.  This
difference in behavior could be an indicator of a BH nature for Ho~IX~X-1, and
perhaps by extension, IC~342 X-1 and M33 X-8.  Although our NS baseline model
was motivated to test whether an accretion column model from W18a was
appropriate for NSs was acceptable, this model can easily be adapted to BHs by
replacing the accretion column with an optically thin disk wind.

Unfortunately, we are unable to determine the nature of the compact object in
M33 X-8 in this study.  Part of the difficulty is that, for the NS baseline
model, the Comptonization component is modeled to be quite weak relative to the disk, similar to that observed for high-luminosity Galactic accreting pulsars (Yang \etal\ 2018).  
%The Comptonization component contributes only $\sim$10\% of the 0.3--40~keV flux, compared with the$\sim$50--80\% found in known pulsating ULXs.  
Thus, if M33 X-8 contains a pulsating NS, pulsations would likely be weak and difficult to detect, even
with broad band data, unless the NS were to go into a bright state, similar to
that found in NGC~5907 ULX1.  On the other hand, significant headway can be
made in understanding the nature of the accretion disk and Comptonization
components by obtaining additional broad band spectra of X-8 in various flux
states.
When using the archival \xmm\ data to interpret changes in the
spectrum, we find degeneracies between the BH and NS models, which could be
broken using data above 10~keV.  In particular, future \xmm\ and \nustar\
observations of X-8 in low ($F_{\rm X} < 1.5 \times 10^{-11}$~\flux) and high
($F_{\rm X} > 1.7 \times 10^{-11}$~\flux) states would allow these degeneracies
to be addressed better.

\acknowledgments L.A.W. and B.D.L. gratefully acknowledge financial support from NASA grant 80NSSC18K039, related to the \xmm\ observations.  We thank the \nustar\ science organizing committee for executing the \nustar\ Legacy Program observations in M33, which were critical to this study.  This research has made use of the NASA/IPAC Extragalactic Database (NED) which is operated by the Jet Propulsion Laboratory, California Institute of Technology, under contract with NASA. %\added{
We thank the referee for their helpful suggestions, which helped improve the quality of this paper.

%\added{
\software{NuSTARDAS (v1.7.1), HEAsoft (v6.20), SAS (v16.0.0; Gabriel et al. 2004), XSPEC (v12.9.1; Arnaud 1996)}



\appendix

\section{Supplemental Fits to Nearly Simultaneous \nustar\ plus \xmm\ Observation}

As an extension to the spectral fits presented in $\S$3.1, Table~A1 provides
parameters and goodness of fit information for four alternative model fits to
the nearly simultaneous \nustar\ plus \xmm\ observations.

%%%%%%%%%%%%%%%%%%%%%%%%%%%%%%%%%%%%%%%%%%%%%%%%%%%%%%%%%%%%%%%%%%%%%%%%%%%%%%%%%%
% Table A1
%%%%%%%%%%%%%%%%%%%%%%%%%%%%%%%%%%%%%%%%%%%%%%%%%%%%%%%%%%%%%%%%%%%%%%%%%%%%%%%%%%
\begin{table}
\renewcommand\thetable{A1}
\begin{center}
\caption{Supplemental Best fit parameters for Nearly-Simultaneous \nustar\ plus \xmm\ Observation.}
\begin{tabular}{lcc}
\hline\hline
\multicolumn{1}{c}{\sc Parameter} & {\sc Unit} & {\sc Value} \\ %
\hline
\multicolumn{3}{c}{{\ttfamily TBABS $\times$ POWERLAW}} \\
\hline
$N_{\rm H}$ \dotfill                           &   10$^{22}$ cm$^{-2}$     & 0.28        \\
$\Gamma$ \dotfill                              &                           & 2.49  \\
$N_{\rm pow}$ \dotfill                                   &   keV$^{-1}$~cm$^{-2}$~s$^{-1}$ at 1 keV      & $6.8 \times 10^{-3}$  \\
$\chi^2/\nu (\nu)$                             &                           & 3.63 (1715)  \\
Null $P$                                       &                           & 0  \\
\hline
\multicolumn{3}{c}{ {\ttfamily TBABS $\times$ BKNPOWER}} \\
\hline
$N_{\rm H}$ \dotfill                           &   10$^{22}$ cm$^{-2}$     & $0.185 \pm 0.005$        \\
$\Gamma_1$ \dotfill                            &                           & $1.97 \pm 0.02$  \\
$E_{\rm break}$ \dotfill                       &                           & $3.70 \pm 0.08$  \\
$\Gamma_2$ \dotfill                            &                           & $3.40 \pm 0.04$  \\
$N_{\rm bknpow}$ \dotfill                                   &   keV$^{-1}$~cm$^{-2}$~s$^{-1}$ at 1 keV      & $(4.77 \pm 0.08) \times 10^{-3}$  \\
$\chi^2/\nu (\nu)$                             &                           & 1.08 (1713)  \\
Null $P$                                       &                           & 0.01  \\
\hline
\multicolumn{3}{c}{{\ttfamily TBABS $\times$ (DISKBB + POWERLAW)}} \\
\hline
$N_{\rm H}$ \dotfill                           &   10$^{22}$ cm$^{-2}$     & $0.207 \pm 0.007$        \\
$kT_{\rm in}$ \dotfill                         &   keV                     & $1.21 \pm 0.02$ \\
$(R_{\rm in}/D_{10})^2 \cos \theta$ \dotfill   &                           & $0.22 \pm 0.02$  \\
$\Gamma$ \dotfill                              &                           & $2.60 \pm 0.03$  \\
$N_{\rm pow}$ \dotfill                         &   keV$^{-1}$~cm$^{-2}$~s$^{-1}$ at 1 keV      & $(3.3 \pm 0.2) \times 10^{-3}$  \\
$\chi^2/\nu (\nu)$                             &                           & 1.01 (1713)  \\
Null $P$                                       &                           & 0.33  \\
\hline
\multicolumn{3}{c}{{\ttfamily TBABS $\times$ (DISKBB + CUTOFFPL)}} \\
\hline
$N_{\rm H}$ \dotfill                           &   10$^{22}$ cm$^{-2}$     & $0.167 \pm 0.008$        \\
$kT_{\rm in}$ \dotfill                         &   keV                     & $1.14 \pm 0.03$ \\
$(R_{\rm in}/D_{10})^2 \cos \theta$ \dotfill   &                           & $0.24_{-0.02}^{+0.03}$  \\
$\Gamma$ \dotfill                              &                           & $2.15 \pm 0.13$  \\
High-$E$ Cut                                   &   keV                     & $11.8_{-2.8}^{+4.9}$ \\
$N_{\rm cuttoffpl}$ \dotfill                   &   keV$^{-1}$~cm$^{-2}$~s$^{-1}$ at 1 keV      & $(3.2 \pm 0.2) \times 10^{-3}$  \\
$\chi^2/\nu (\nu)$                             &                           & 0.99 (1712)  \\
Null $P$                                       &                           & 0.56  \\
\hline
\hline
\end{tabular}
\end{center}
Note---All quoted errors are at the 90\% confidence level.  Given the very
poor fit provided by the {\ttfamily TBABS $\times$ POWERLAW} model, errors are
not reported.\\
\end{table}


\end{document}